\documentclass[12pt,preprint]{aastex}

\begin{document}


\title{Spectral Variability of Quasars in the Sloan Digital Sky Survey. I: Wavelength Dependence}


\author{
Brian C. Wilhite\altaffilmark{1,2},
Daniel E. Vanden Berk\altaffilmark{3,4},
Richard G. Kron\altaffilmark{1,5},
Donald P. Schneider\altaffilmark{4},
Nicholas Pereyra\altaffilmark{3},
Robert J. Brunner\altaffilmark{6,7},
Gordon T. Richards\altaffilmark{8},
Jonathan V. Brinkmann\altaffilmark{9}
}

\altaffiltext{1}{The University of Chicago, Department of Astronomy
  and Astrophysics, 5640 S. Ellis Ave., Chicago, IL 60637}
\altaffiltext{2}{Current address: University of Illinois, Department of Astronomy, 1002 W. Green St., Urbana, IL 61801; wilhite@astro.uiuc.edu}
\altaffiltext{3}{The University of Pittsburgh, Department of Physics and Astronomy, Pittsburgh, PA 15260}
\altaffiltext{4}{The Pennsylvania State University, Department of Astronomy and Astrophysics, 525 Davey Lab, University Park, PA 16802}
\altaffiltext{5}{Fermi National Accelerator Laboratory, P.O. Box 500,
  Batavia, IL 60510}
\altaffiltext{6}{The University of Illinois, Department of Astronomy,
  1002 W. Green St., Urbana, IL 61801}
\altaffiltext{7}{National Center for Supercomputing Applications, 605 E. Springfield Ave., Champaign, IL 61820}
\altaffiltext{8}{Princeton University Observatory, Peyton Hall, Princeton, NJ 08544}
\altaffiltext{9}{Apache Point Observatory, P.O. Box 59, Sunspot, NM 88349}


\begin{abstract}
Sloan Digital Sky Survey (SDSS) repeat spectroscopic observations have
resulted in multiple-epoch spectroscopy for $\sim 2500$ quasars observed more than 50 days apart. 
From this sample, calibrating against stars observed simultaneously, we identify 315 quasars that have varied significantly 
between observations (with respect to assumed 
non-variable stars observed concurrently).  These variable quasars range in redshift from 0.5 to 4.72.
This is the first large quasar sample studied spectroscopically for variability and represents a potentially 
useful sample for future high-redshift reverberation mapping studies.
This also marks the first time the precise wavelength dependence of quasar variability has been determined, allowing both the 
continuum and emission line variability to be studied.  We create an ensemble 
difference spectrum (bright phase minus faint phase) covering rest-frame wavelengths from 1000\AA\ to 6000\AA.
This average difference spectrum is bluer than the average single-epoch quasar spectrum; a power-law fit to the
difference spectrum yields a spectral index $\alpha_{\lambda} = -2.00$, compared to an index of $\alpha_{\lambda} = -1.35$ for 
the single-epoch spectrum.  This confirms that quasar continua are bluer when brighter.  The difference spectrum also exhibits very weak or absent emission line features; the strongest emission lines vary only $30\%$ as much as the continuum.  This small emission line variability with respect to the continuum is consistent with the Intrinsic Baldwin Effect.  Due to the lack of variability of the lines, measured photometric color is not always bluer in brighter phases, but depends on redshift and the filters used.
Lastly, the difference spectrum is bluer than the ensemble quasar
spectrum only for $\lambda_{rest} < 2500$\AA, indicating that the variability cannot
result from a simple scaling of the average quasar spectrum.
\end{abstract}



\section{Introduction\label{intro}}

The luminosities of quasars and other active galactic nuclei
(AGN) have been observed to vary on time scales from hours to decades, 
and from X-ray to radio wavelengths.
The majority of quasars exhibit continuum variability on the order of 
$20\%$ on timescales of months to years \citep{hook94}.  
In fact, variability has long been used as a selection criterion in creating 
quasar samples from photometric data \citep[e.g.,][]{ivezic04b, rengstorf04, koo86}.

Many simple correlations between photometric variability and various physical 
parameters have been known for decades.  These relationships are 
summarized in \citet{helfand01} and \citet{giveon99}.  Numerous studies \citep[e.g.,][]{hawkins02,devries03} have shown variability to 
correlate with time lag.  Anti-correlations have been found between 
variability and luminosity \citep[e.g.,][]{uomoto76,cristiani97}
and wavelength \citep[e.g.,][] {giveon99, trevese01}.   
Recently, \citet[hereafter VB04]{vandenberk04}, using a sample of $\sim25000$ quasars, confirmed these known 
correlations, and parameterized relationships between variability 
and time lag, luminosity, rest-frame wavelength and redshift.

To this point, work on quasar spectral variability has focused on individual 
objects.  Reverberation mapping studies have included hundreds of observations 
of $\sim30$ objects, with the primary goals of estimating the central black hole 
mass and gaining an understanding of the structure of the broad-line emitting 
region \citep{peterson93}.  As low-luminosity 
AGN are known to be more variable, reverberation mapping studies have 
concentrated on nearby, low-luminosity Seyfert I galaxies.
To date, there has been no study of spectral variability of an entire ensemble of quasars.  
Although several studies have shown a strong anti-correlation between wavelength 
and variability, VB04 was the first to parameterize quasar variability 
as a function of wavelength.  In that paper, variability 
was measured at $\sim 100$ values of wavelength.  Spectroscopy has the potential 
for a far more detailed measurement of variability versus wavelength.

In this paper we present results on a quasar spectral variability program using
data from the Sloan Digital Sky Survey \citep[SDSS;][]{york00}.  This paper builds 
on the spectrophotometric calibrations of VB04, this time comparing 
stars and quasars at multiple spectroscopic epochs. 

We describe the quasar sample, the additional necessary photometric 
calibrations, and the identification of variable quasars in \S\,\ref{dataset}.  
We describe the calculation of composite difference spectra in 
\S\,\ref{composite}.  The results are discussed in 
\S\,\ref{discussion},  and we conclude in \S\,\ref{conclusions}.

Throughout the paper we assume a flat, cosmological-constant-dominated cosmology 
with parameter values $\Omega_\Lambda = 0.7, \Omega_{M} = 0.3,$ and $H_{0}=70$km s$^{-1}$ Mpc$^{-1}$.


\section{The Quasar Dataset\label{dataset}}

%
%
\subsection{The Sloan Digital Sky Survey\label{SDSS}}
Through Summer 2004, the Sloan Digital Sky Survey \citep[SDSS;][]{york00} has imaged almost $\sim8200$ deg$^{2}$ and obtained follow-up spectra for roughly $5 \times 10^{5}$ galaxies and $5 \times 10^{4}$ quasars.
All observations are made at the Apache Point Observatory 
in New Mexico, using a dedicated 2.5-m telescope.  Imaging is done with a
mosaic CCD camera which operates in drift-scanning mode \citep{gunn98}.  Absolute 
astrometry for point sources is accurate to better than $100$ milliarcseconds 
\citep{pier03}.  Site photometricity and extinction monitoring
are carried out simultaneously with a 0.5m telescope at
the observing site \citep{hogg01}.  Imaging data are reduced and calibrated
using the {\tt PHOTO} software pipeline \citep{lupton01}.  Imaging quality control is discussed in \citet{ivezic04a}.


%
%
\subsection{Quasar Target Selection and Sample Definition\label{sample}}

Objects in the imaging survey are selected for spectroscopic follow-up 
as candidate galaxies \citep{strauss02,eisenstein01}, quasars 
\citep{richards02}, and stars \citep{stoughton02} and grouped by 3-degree diameter 
areas or ``tiles'' \citep{blanton03a}.  For each tile, an aluminum plate is 
drilled with hole positions corresponding to the sky locations of the targets.  
Plates are placed at the focal plane of the telescope, 
where the holes subtend a 3-arcsecond diameter.  Optical fibers are plugged into 
the holes and run from the plate to one of two spectrographs, each of which 
accepts 320 fibers.  This allows for the simultaneous observation of 640 objects.  On 
each plate, approximately 500 galaxies, 50 quasars and 50 stars are observed (the 
remaining fibers are reserved for blank sky and calibration stars).  
Spectroscopic 
observations generally occur up to a few months, but occasionally years, after the 
corresponding imaging observations, depending upon scheduling constraints.

\begin{figure}
\plotone{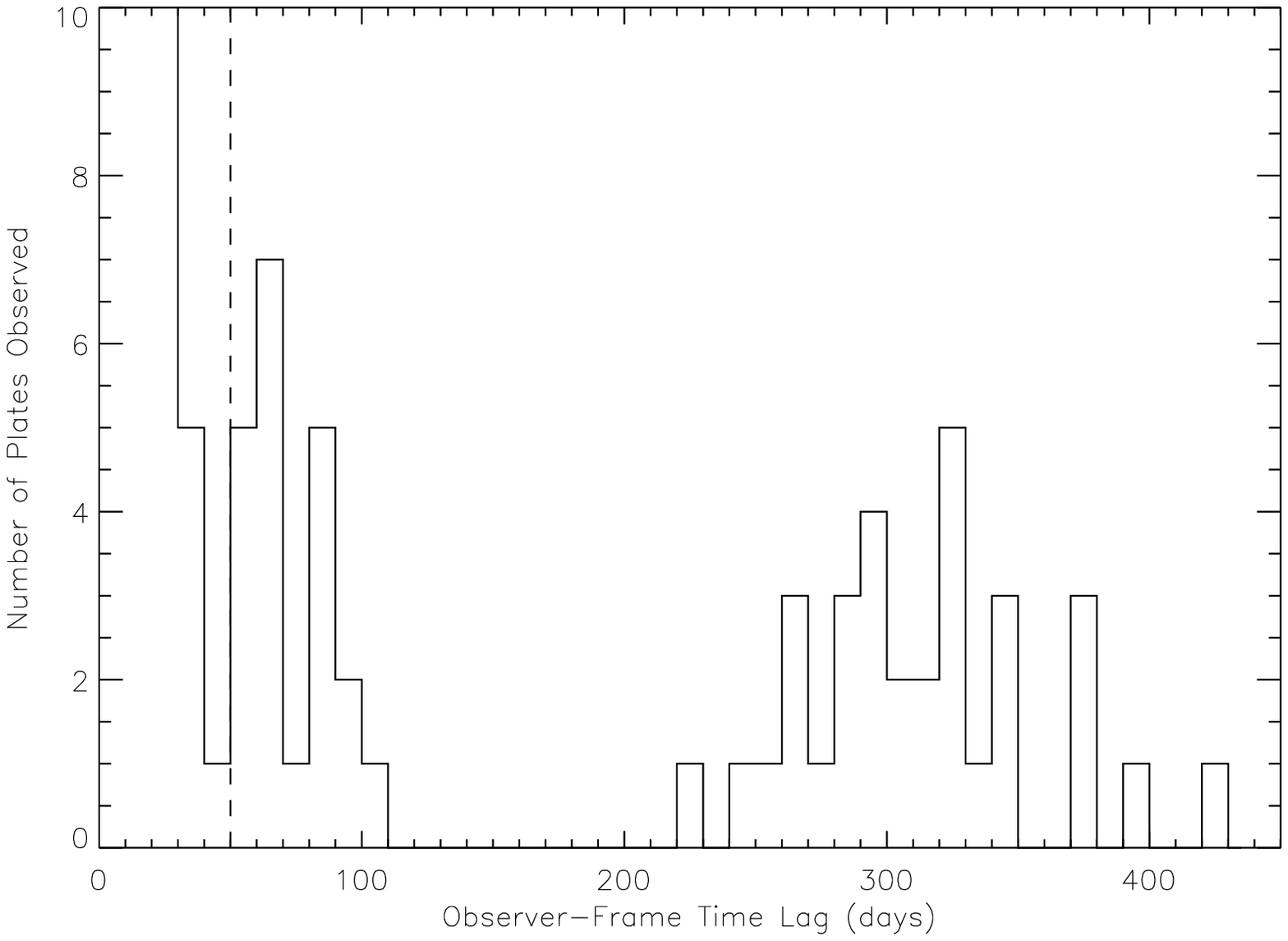}
\caption{Number of plate pairs observed as a function of time between observations.  Plates with time lags greater than 50 days were re-plugged and, therefore, their spectra were not co-added.  Only these large-time-lag plates, representing 53 of the 181 total, are used in this paper.
\label{Fig2.1}}
\end{figure}

Throughout the course of normal operations, the SDSS has taken 
multiple observations of 181 spectroscopic plates through 11 June
2004.  As seen in Figure \ref{Fig2.1}, these observations were 
made with time lags ranging from days to more than one year.  
Repeat spectroscopic observations are taken for a number of reasons.
For example, early in the survey, a reliable method for determining when a plate had 
attained sufficient signal-to-noise ratio had not yet been developed, 
and some early plates that were  satisfactory were re-observed.  Additionally, 
some plates were re-observed by design (i.e. observing conditions were not photometric 
and no unobserved plates were yet available).

When the time separation between observations is small (i.e. less than a month), spectra from multiple nights'
observations are co-added as part of the {\tt Spectro2d} processing pipeline to increase the spectral signal-to-noise ratio.  
Only nine cartridges exist with which to mount plugged plates on the telescope.  
As only a finite number of plates may remain plugged 
at any given time, plates which are not scheduled for observation (or 
re-observation) in the next $\sim30$ days have their fibers removed and are set aside for potential later use.  
Thus, when observations are separated by more than 
a month, plates must be re-plugged---the spectroscopic fibers are re-fitted 
into the 640 holes on a plate, with no attempt made to plug them into the holes they 
originally occupied.  In these cases, the pipeline software does not co-add spectra from different dates of observation.  
We use only plates whose 
observations were separated by more than 
50 days, to ensure that none of our spectra are co-additions of spectra from multiple nights.  As quasar variability is known to increase with time lag, 
these largest time-lag plates are most relevant to this study.  Two large-time-lag plates are not 
used as they do not contain any useful quasars ($z > 0.5$; see \S\,\ref{spectro}).
This leaves 53 spectroscopic plate pairs for study, containing a total of roughly 3000 stars and 2500 quasars.
In the rare cases where a plate is observed more than two times, we use only the 
first and last observations of that plate, in order to ensure the longest possible time lag.
Single observations of 34 of the 53 large-time-lag plate pairs have 
been publicly available since April 2003 as part of the SDSS First Data 
Release \citep[DR1;][]{abazajian03}.  Observations of eight of the remaining plates were released in 
March 2004 with the Second Data Release \citep[DR2][]{abazajian04}.  Spectra from six other plates were released as part of the SDSS Third Data Release \citep[DR3;][]{abazajian05}.
A list of plates used, and the corresponding data releases, can be found in Table \ref{platetable}.

Quasar candidates are selected from the imaging sample by their
non-stellar colors from the five-band photometry as well as by matching
SDSS point sources with FIRST radio sources \citep{richards02}.
About two-thirds of the candidates are confirmed to be quasars from
the spectroscopic survey.  Ultraviolet excess quasars are targeted to
a limit of $i = 19.1$ and higher redshift quasars are targeted to $i =
20.2$.  The completeness of the quasar sample selected by the SDSS algorithm is 
approximately 95\% for unresolved sources to the $i$=19.1 magnitude limit \citep{vandenberk05}.
Additional quasars are targeted as part of the SERENDIPITY
and ROSAT classes \citep{stoughton02} or (incorrectly) as stars.  Our initial sample
includes objects selected in any of these ways.

%
%
\subsection{SDSS Spectroscopy and Its Calibration\label{spectro}}

Spectra are obtained in at least three exposures of typically 15 minutes' duration.
There are 32 sky fibers, 8 spectrophotometric
standard stars and 8 reddening standard stars observed on a typical
plate to help with calibration of the remaining 592 science targets.
Spectral reductions and calibrations are done using the SDSS {\tt Spectro2d}
pipeline \citep{stoughton02}.  The 8 spectrophotometric calibration
stars are chosen to approximate the standard F0 subdwarf star $BD+17\degr4708$,
and are used by the {\tt Spectro2d} pipeline for absolute spectral flux calibration.
Unlike those spectra released in the EDR and DR1, the spectra used here are not 
corrected for reddening due to Galactic extinction.  The most recent version of the 
spectroscopic pipeline (version 23) introduced appreciable improvements in spectrophotometry,
but does not correct for reddening \citep{abazajian04}.  
The average correction in DR2 is quite small; $<E(B-V)> = 0.034$ \citep{abazajian04}.  
In principle, this should have no effect on the selection of variable objects, as the same 
reddening law should apply at both epochs.  Thus, we do not make a reddening correction here.
Composite difference and single-epoch quasar spectra (see \S\,\ref{composite}) will remain slightly reddened, but ratios of these spectra will not.
SDSS spectra cover 3900\AA\ to 9100\AA\ and have a resolution of $\lambda/\Delta{\lambda} = 
1800-2100$, corresponding to a bin width of less than 3\AA\ at 5000\AA.  
Spectra are extracted in equal bins in log($\lambda$).  
Therefore, spectral features of the same velocity width cover the same number of re-binned pixels, regardless of redshift.


A separate processing pipeline called {\tt Spectro1d} does line identification and 
measurement, redshift determination, and spectral classification.
Quasars are identified from their spectra using a combination of both
automated classification (about $94\%$) and manual inspection
of those objects flagged by the spectroscopic pipeline as being
less reliably identified (about $6\%$).  For the purposes of this study, we define
``quasar'' to mean any extragalactic object with broad emission lines
(full width at half maximum velocity width of $\gtrsim 1000$ km s$^{-1}$),
regardless of luminosity.  The definition thus includes objects which
are often classified as less luminous types of active galactic nuclei
rather than quasars, and excludes AGN without strong broad
emission lines such as BL~Lacs and some extreme broad absorption
line quasars.  

Stars are identified in spectroscopy by cross-correlating 
their spectra with one of 15 stellar templates.  A few percent of the objects ultimately 
classified as stars are identified through manual inspection \citep{stoughton02}.  
To assemble our sample, we utilize spectra for all objects 
spectroscopically confirmed as stars or quasars.

The {\tt Spectro1d} pipeline flags potentially problematic pixels for reasons
ranging from inadequate sky subtraction to cosmic ray rejection.  We use those flags that imply problems that could adversely affect spectrophotometry: 
SP\_BADTRACE, SP\_BADFLAT, SP\_BADARC, SP\_NEARBADPIX, SP\_LOWFLAT, SP\_FULLREJECT, SP\_SCATLIGHT, SP\_BRIGHTSKY, SP\_NODATA and SP\_COMBINEREJ.  A full description of these (and other) flags is given by Stoughton \citep{stoughton02}.
Pixels with any one of the flag values from the above list are rejected from the analysis which follows.  Any 
object (star or quasar) with more than $50\%$ of its pixels flagged is removed 
from its respective sample and not used in this work. This removes fewer than 
one star and quasar per plate.

Ideally, in low-redshift AGN, the underlying galaxy spectrum would be 
easily removable when comparing spectra from different 
epochs.  However, due to the extended nature of galaxy images, 
a small difference in pointing or seeing between epochs can cause a 
large enough change in flux to cause the appearance of variability.  To avoid 
including these potential false positives, we include only those quasars with 
redshifts greater than 0.5.  This leaves 2181 objects.



%
%
\subsection{Refinement of Spectroscopic Calibration \label{calib}}

In spectral variability studies, accurate spectrophotometry is essential.  As was 
shown in VB04, some additional calibration of SDSS spectrophotometry 
is necessary to achieve the errors required to fully 
extract all AGN variability information.  Here, as in VB04, the stars corresponding to an individual plate 
are used to calibrate the flux of all point sources observed with that plate, under the 
assumption that the stars are a non-variable population (evident variables 
are removed from the calibration, as discussed later).

For SDSS spectra, the signal-to-noise ratio (S/N) in each pixel is the ratio of the flux 
to the error as determined by {\tt Spectro2d} \citep{stoughton02}.  {\tt Spectro1d} determines three values of 
S/N for a spectrum by calculating the median S/N in the pixels corresponding to wavelengths covered by the SDSS $g, r$ and $i$ filter transmission curves.
Hereafter, when referring to the two halves of a plate 
pair, we use ``high-S/N'' for the plate with the 
higher median $r$-band stellar signal-to-noise ratio.  The plate 
with the lower median $r$-band stellar signal-to-noise ratio will be called ``low-S/N.''  
This is a plate-wide designation; 
it does not speak to the relative S/N values for any given individual object.
The lower signal-to-noise ratio epoch stellar spectra will be scaled to match the average 
flux of the high-S/N epoch stellar spectra to correct for epoch-to-epoch spectrophotometric 
calibration differences.  

\begin{figure}
\plotone{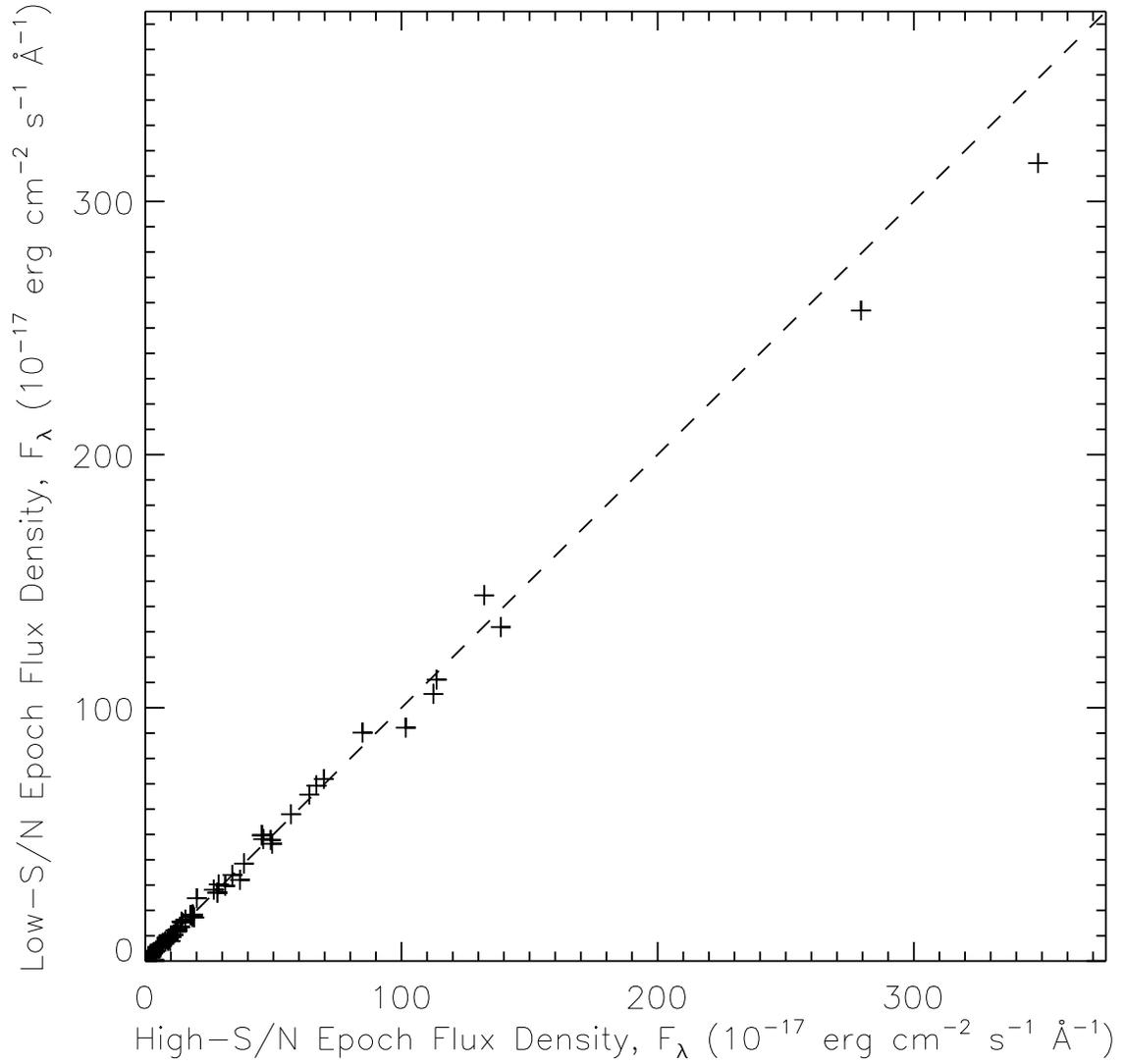}
\caption{The flux density at the high-S/N epoch versus the flux density at the low-S/N epoch at 5000\AA\ for 71 calibration stars on plate 547.  The dashed line represents $F_{HSN}=F_{LSN}$.
\label{Fig2.2}}
\end{figure}

To begin calibration, we take the median ratio of the flux in a pixel at the high-S/N 
to the flux in that same pixel at the low-S/N epoch for all $\sim 50$ stars
on a plate (see Fig.\,\ref{Fig2.2}).  This median ratio represents the correction 
value at that wavelength.  Computing 
this value for all wavelengths gives the initial correction 
spectrum for that plate.  In the case of perfect spectrophotometric calibration 
and a completely non-variable stellar population, this spectrum would have a value 
of 1 at all wavelengths.  To scale the lower signal-to-noise-ratio epoch observations, 
all low-S/N epoch stellar spectra are multiplied by the initial correction spectrum for 
the corresponding plate.

By using the stars to calibrate relative fluxes, we are making the assumption that 
the stellar population is non-variable.  In truth, some fraction of the stars will 
be variable objects that need to be excluded from the calibration as they may skew 
the final correction spectrum.  To measure the size of the stellar flux variations, we use 
the integrated relative flux change, ($\Delta f/f$).  For each star, we sum over 
high-S/N and low-S/N epoch fluxes (yielding $f_{HSN}$ and $f_{LSN}$).  We then take the 
ratio of the difference ($f_{HSN}-f_{LSN} = \Delta f$) to the average ($\frac{f_{HSN}+f_{LSN}}{2} 
= f$) which gives a measure of the total relative flux change between epochs.  

\begin{figure}
\plotone{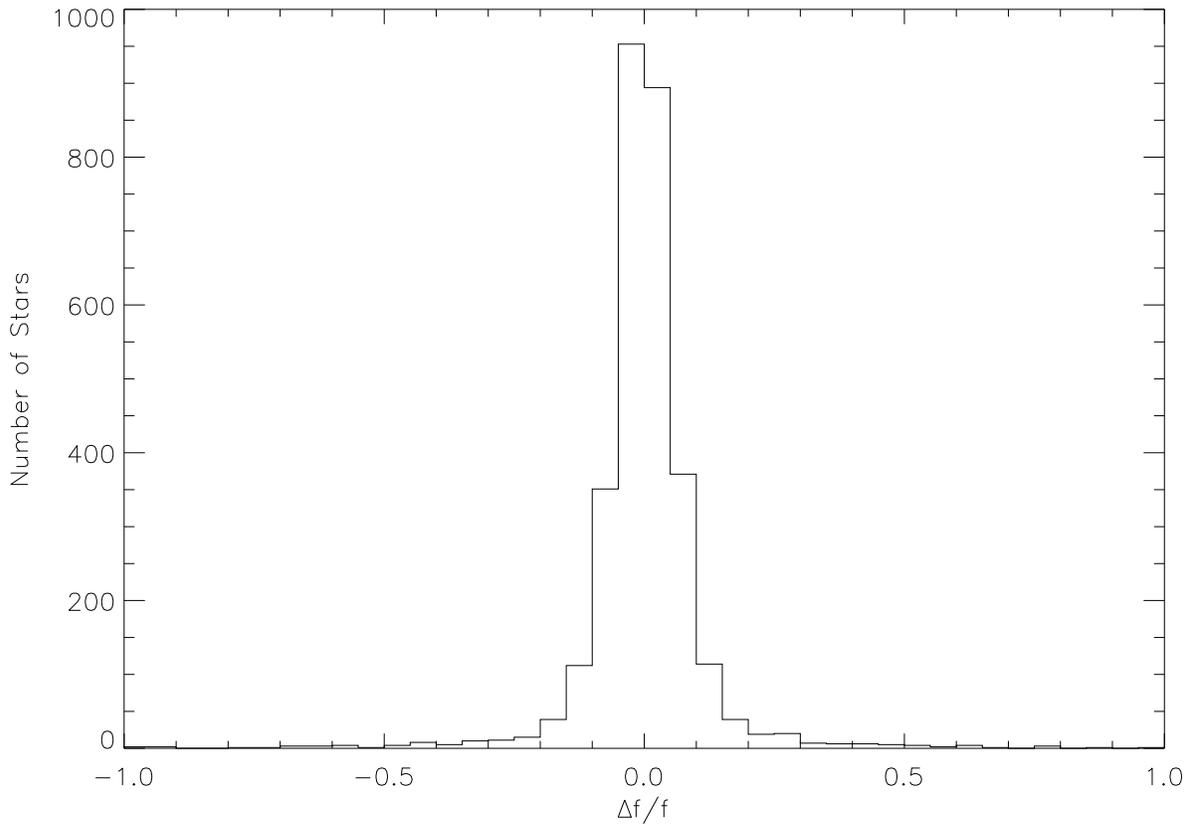}
\caption{Integrated relative flux change ($\Delta f/f$) for all stars on the 53 plates with more than 50 days time lag between observations.  The overpopulation of the wings of the distribution is a sign that the excluded stars are truly variable.
\label{Fig2.2.5}}
\end{figure}

That some of these stars is variable is seen in the histogram of stellar $\Delta f/f$ values in Fig. \ref{Fig2.2.5}.
While the center of the distribution is Gaussian, the tails are overpopulated, relative to a Gaussian distribution, a clear indication that the largest flux variations are not due to simple shot noise.

\begin{figure}
\plotone{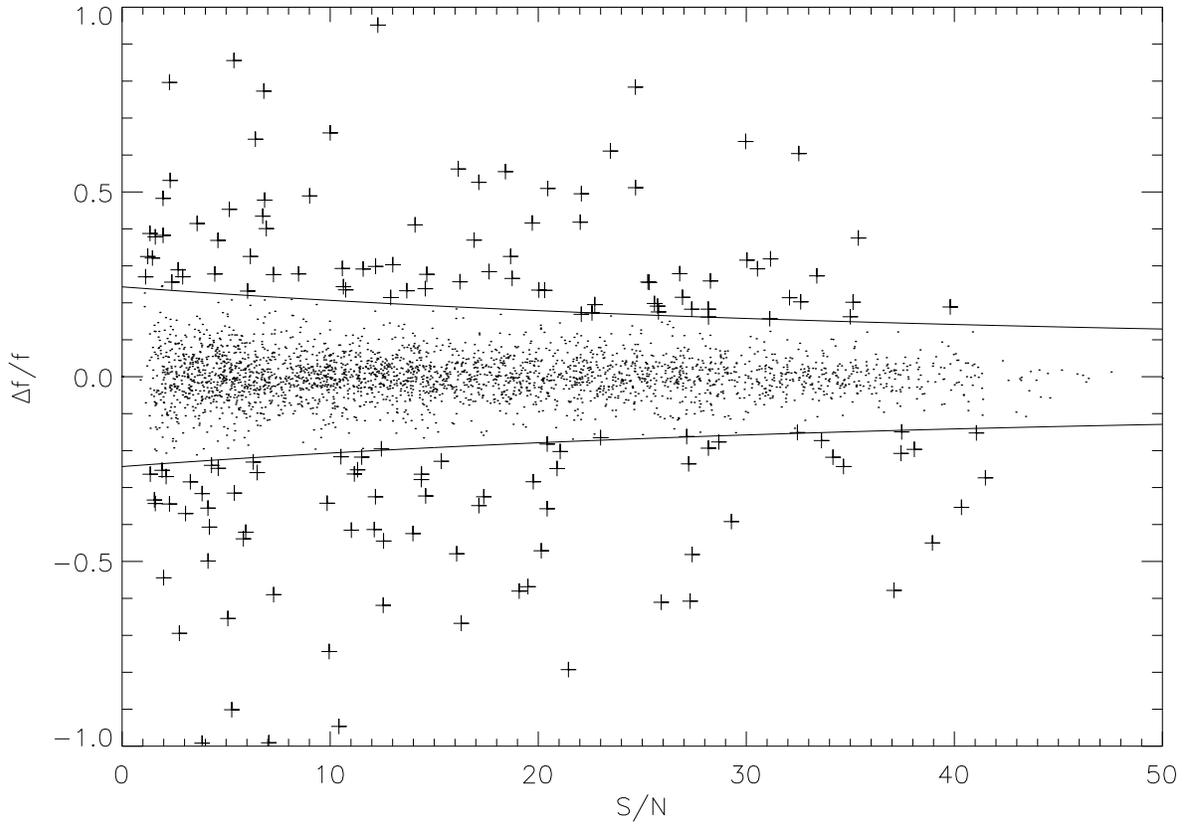}
\caption{Integrated relative flux change ($\Delta f/f$) versus high-S/N epoch signal-to-noise ratio (S/N) for all stars on the 53 plates with more than 50 days time lag between observations.  The overlaid curves show three times the $68.3\%$ confidence interval of $\Delta f/f$ for given values of S/N for all stars.  Those stars outside these envelopes (marked with crosses) are assumed to be variable and are not used in re-calibration.
\label{Fig2.3}}
\end{figure}

As seen in Figure 2 of VB04, the width of the stellar relative flux change distribution is a strong 
function of spectral signal-to-noise ratio.  Thus, to determine which stars are likely to be variable, we 
must account for these S/N dependences.  
We calculate an average signal-to-noise ratio for every star and quasar by taking the ratio of the 
average inegrated flux ($f$, from above) to the integrated 
sum in quadrature of the errors at the two epochs ($\sqrt{\sigma_{HSN}^{2}+\sigma_{LSN}^{2}} = \sigma$).  
We plot $\Delta f/f$ versus the high-S/N epoch signal-to-noise for all stars on all plates in Fig. \ref{Fig2.3}. 

To characterize the width of the $\Delta f/f$ distribution as a function of S/N, we separate the 
stars into 10 bins in signal-to-noise ratio and calculate the $68.3\%$ 
confidence interval in $\Delta{f/f}$ in each bin.  We then take the average of the 
upper boundary and the absolute value of the lower boundary of the interval.  
To this average, we fit an exponential envelope:

\begin{equation}
E(S/N) = a_{0}e^{{\frac{S/N}{a_{1}}}} + a_{2}.  
\end{equation}

The 177 stars ($5.8\%$ of the total of 3074) with $|\Delta{f/f}|$
greater than three times this envelope are rejected from the calibration sample, 
under the assumption that they are variable (these are marked with crosses in Fig. \ref{Fig2.3}).  
This gives an average of roughly 3.5 variable stars 
removed per plate, with a maximum of 16 variable stars (out of 64 total stars) removed from plate 678.  

\begin{figure}
\plotone{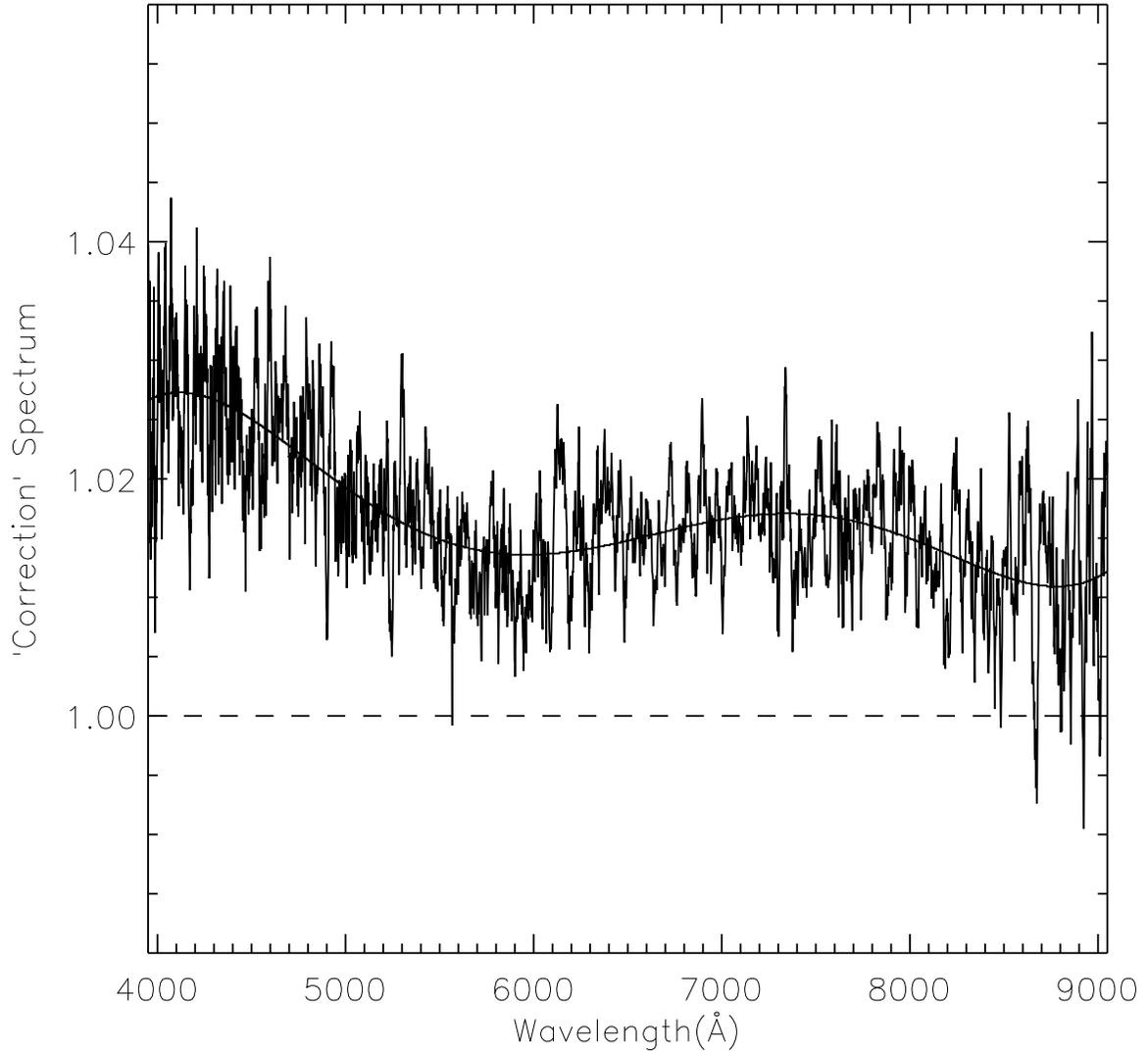}
\caption{Final ``correction'' spectrum, derived from 71 stars, for plate 547.  All low-S/N epoch quasar spectra on plate 547 are multiplied by the fifth-order polynomial fit seen here to match the spectrophotometric calibration of the high-S/N epoch.  Note that the largest corrections are only of the order of a few percent. 
\label{Fig2.4}}
\end{figure}

After the variable stars are removed, we use the ratio of the high-S/N and low-S/N epoch 
fluxes for all remaining stars on a plate to compute the final correction spectrum.  
To remove the noise in the correction spectrum due to shot noise, we fit a fifth-order 
polynomial to use as the final correction spectrum (see Fig. \ref{Fig2.4}).
This re-scaling is applied to all low-S/N epoch quasar and stellar spectra on the plate (as well as the the low-S/N epoch spectral error). 

\begin{figure}
\plotone{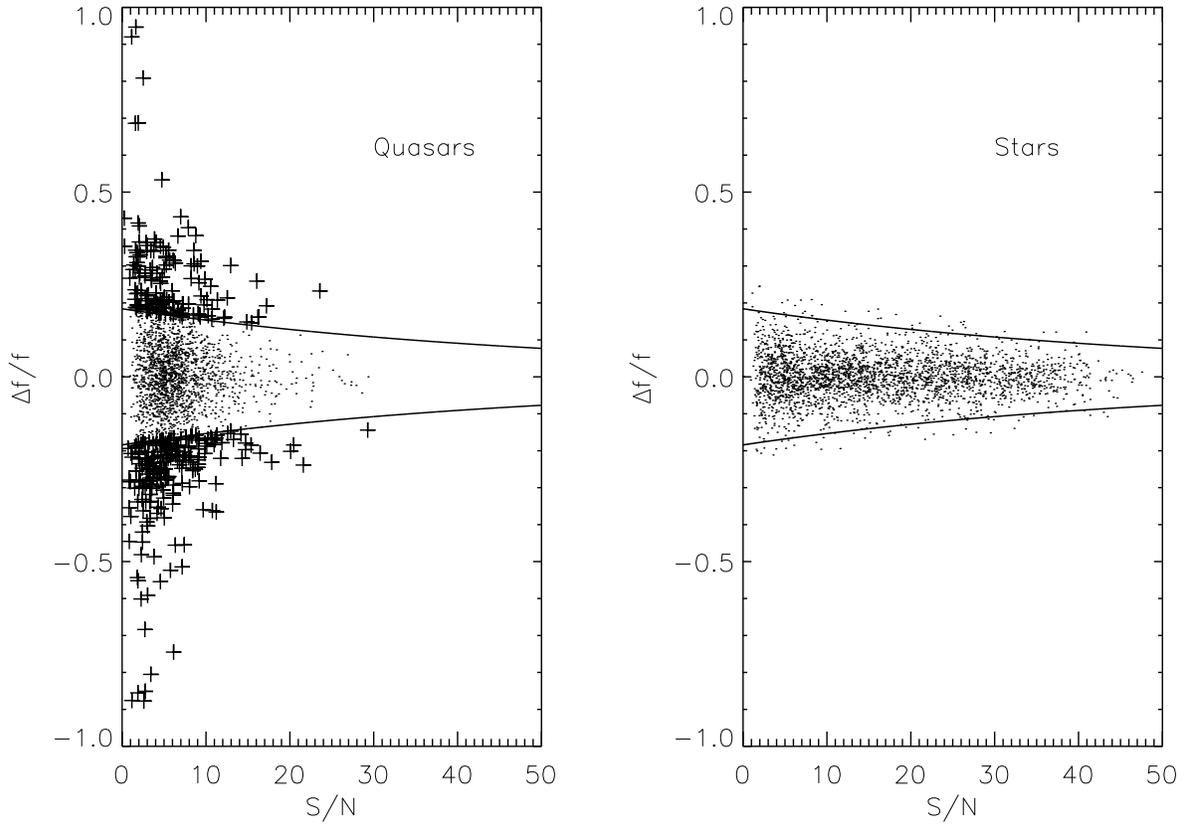}
\caption{Integrated relative flux change ($\Delta{f/f}$) versus high-S/N epoch signal-to-noise ratio (S/N) for quasars (left) and stars (right) on all 53 plates with more than 50 days time lag between observations.  The overlaid curves show 2.5 times the $68.3\%$ confidence interval of $\Delta{f/f}$ for given values of S/N for all stars.  The 364 quasars marked with crosses are selected to be candidates for the variable quasar sample.
\label{Fig2.5}}
\end{figure}

We then re-calculate $f_{HSN}$,$f_{LSN}$,$f$ and $\Delta f/f$ by again integrating over all high-S/N epoch  and newly 
re-scaled low-S/N epoch spectra of both stars and quasars. 
Fig. \ref{Fig2.5} shows integrated 
relative flux change ($\Delta f/f$) versus signal-to-noise ratio for all quasars 
(as well as the non-variable stars, for reference) on all plates.  
To characterize the width of the distribution as a function of S/N, an exponential envelope is again 
fit to the stellar $68.3\%$ confidence interval for 10 
bins in S/N.  As will be discussed below, 2.5 times this envelope is plotted in Fig. \ref{Fig2.5}.

One can quantify how much an individual 
quasar has varied by dividing the absolute value of the integrated 
relative flux change ($|\Delta{f/f}|$) by the value of the exponential envelope from Fig. \ref{Fig2.5}
at the corresponding value of signal-to-noise:

\begin{equation}
V = \frac{|\Delta{f/f}|}{E(S/N)}.
\end{equation}

\begin{figure}
\plotone{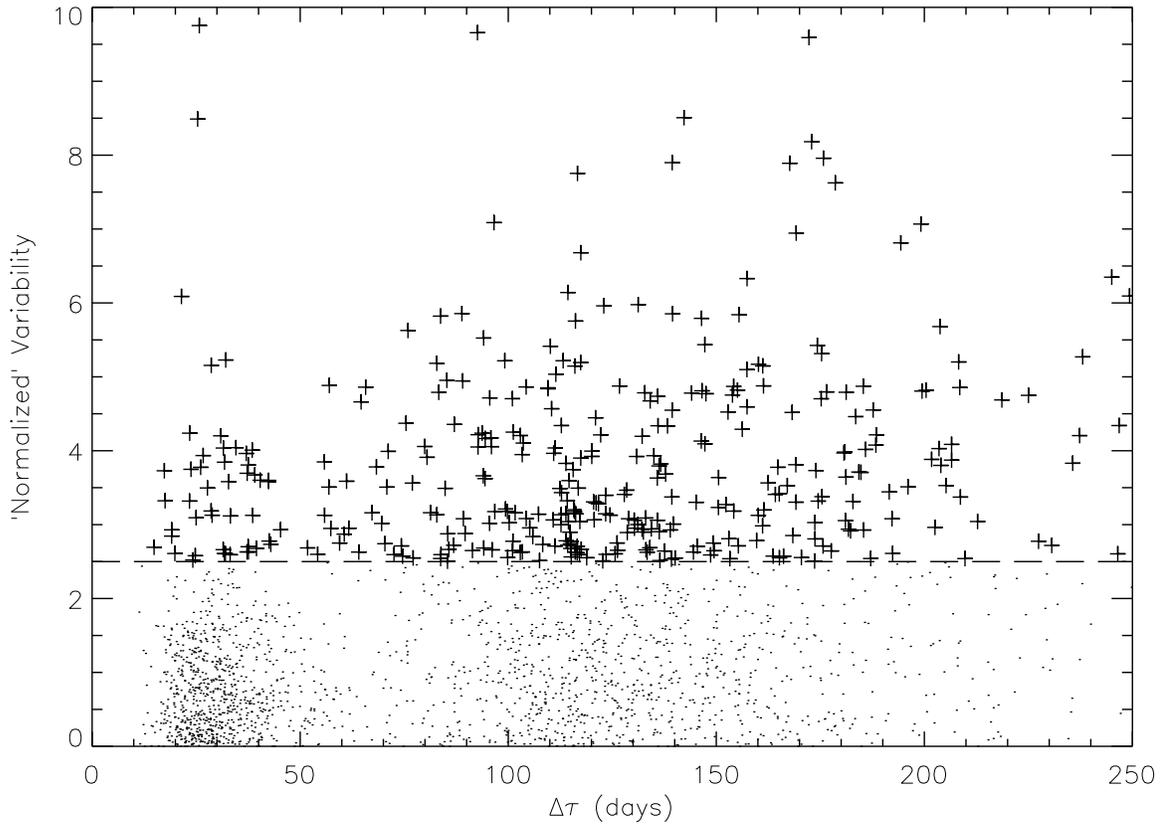}
\caption{Variability (V) versus rest-frame time lag ($\Delta{\tau}$) for 2181 quasars from large time lag plates.  The 364 quasars with values of V greater than 2.5 are selected to be candidates for the variable quasar sample.
\label{Fig2.6}}
\end{figure}

Fig. \ref{Fig2.6} shows this normalized variability ($V$) versus rest-frame time lag ($\Delta{\tau}$) for all 
quasars.  As variability is strongly dependent on rest-frame time lag, it is not surprising that 
most low-$\Delta{\tau}$ ($\Delta{\tau} < 50$ days) quasars are concentrated at low values of $V$.
However, there is a small number of quasars with larger $V$ separated from the rest of the 
low-$\Delta{\tau}$ distribution.  We assume that these are the quasars which have varied significantly.  
We then apply this criterion ($V > 2.5$) to select those variable quasars.  
This yields 364 variable quasar candidates.  Our goal is not to create a complete sample of variable 
quasars, but rather to identify a set of quasars which have varied.

These 364 spectra are inspected manually to ensure that they are quasars with correctly measured redshifts, and that their spectra are free of any serious problems that could affect spectrophotometry.  In a small number of objects (14), either the high-S/N or low-S/N epoch redshift was incorrect.  In these cases, the incorrect redshift was simply changed to match the redshift from the other epoch.  Two objects were removed for weak or virtually nonexistent emission lines, making reliable redshift determinations impossible.  
Three spectra were removed from the sample for demonstrating serious sky-subtraction problems at wavelengths greater than 7000\AA.  It is possible that sky-subtraction differences, not real variability, caused these objects to appear variable.
Thirty objects were removed from the sample for evidence of broad absorption lines.  While there is little evidence
that the variability properties of BALQSOs differ from those of ``normal'' quasars (VB04), the continuum properties are clearly
different \citep[e.g.,][]{reichard03,brotherton01} and BALQSO emission lines are often difficult to measure due to the strong absorption troughs.
Lastly, all fourteen objects from the second half (fiber number greater than 320) of plate 418 were 
removed.  For unknown reasons, these spectra were clearly of a lower 
quality than those from other plates, and only a few of the spectra resembled quasar spectra.  
After removing these 49 spectra, 315 remained to form the variable quasar sample.

That the quasar sample is, on the whole, more variable than its stellar counterpart is 
easily seen in Fig. \ref{Fig2.5}.  Not only are there more quasars beyond
the exponential envelope (even if variable stars removed earlier were to be included), 
but the entire distribution of $\Delta{f}/f$ is wider.  For stars, the 
standard deviation of the $\Delta{f}/f$ distribution (including variable stars) is 0.151.  For quasars, it is 0.213.

\subsection{Variable Quasar Sample\label{variability}}
The set of 315 quasars that have been determined to vary spectroscopically is defined as the variable quasar sample.  Other quasars (perhaps all) are also likely to be variable, but this could not be directly determined from the two-epoch spectroscopic data.
Synthetic apparent magnitudes were determined from the spectra, by convolving the high-S/N epoch spectra with 
the SDSS filter curves.  Absolute magnitudes in the rest frame 
$r$ band are calculated for each quasar using these apparent magnitudes 
and assuming a power law spectral energy distribution 
$f_\lambda\propto \lambda^{\alpha_{\lambda}}$, with a slope of 
$\alpha_\lambda=-1.5$.  Estimated K-corrections using the composite spectrum 
from \citet{vandenberk01} are consistent with the simple power law assumption, and the
differences are usually no greater than 0.1 magnitude at any redshift.
The members of the variable sample have a median absolute magnitude of $M_{r} = 
-24.9$, near to the median of sample of all multiply observed quasars, $M_{r} = -25.2$.  

The redshift distribution of the variable quasar sample is also indistinguishable from the full quasar sample.
Variable quasars have a median redshift $z = 1.47$ while the median redshift of all quasars
is $z = 1.51$.  The most obvious difference between the variable and non-variable samples 
is in rest-frame time lag.  The median rest-frame time lag for all quasars is 98 days, 
but the variable quasars have a median rest-frame time lag of $\Delta{\tau} = 123$ days.  
Given the known dependence of variability on rest-frame time lag, the difference 
between samples is unsurprising.  

\begin{figure}
\plotone{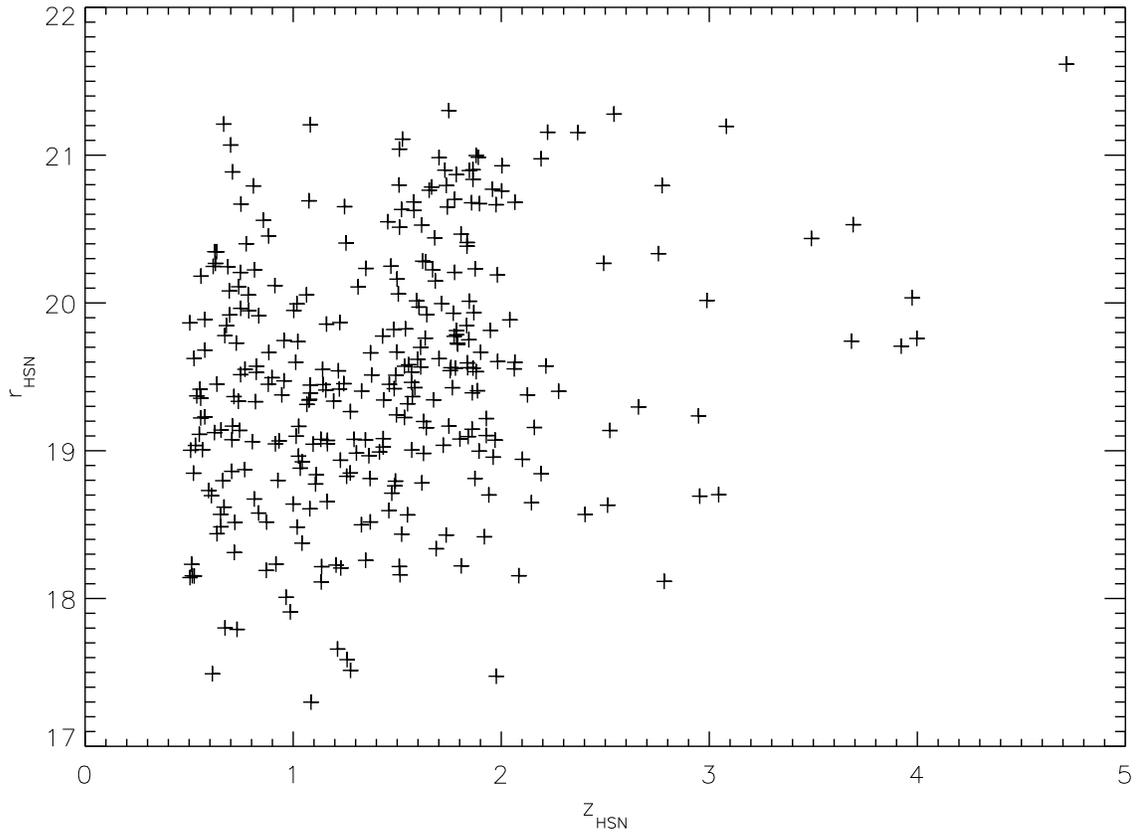}
\caption{High-S/N epoch redshift versus $r$-band apparent magnitude for all 315 objects in the variable quasar sample.
\label{Fig2.6.5}}
\end{figure}

Table \ref{qsotable} lists pertinent information for all members of the sample, including 
SDSS name (which includes right ascension and declination), dates of observation (MJD), redshift (z), apparent ($r$) and absolute ($M_{r}$) $r$-band magnitudes, and $r$-band signal-to-noise ratio ($S/N_{r}$).  Fig. \ref{Fig2.6.5} shows the distribution of these variable quasars in redshift and apparent magnitude.
Spectra for the objects in the variable quasars sample will be published as part of a larger SDSS quasar 
variability catalog, which will contain multi-epoch spectroscopic data for all 2181 quasars on the large time-lag plates.

\begin{figure}
\plotone{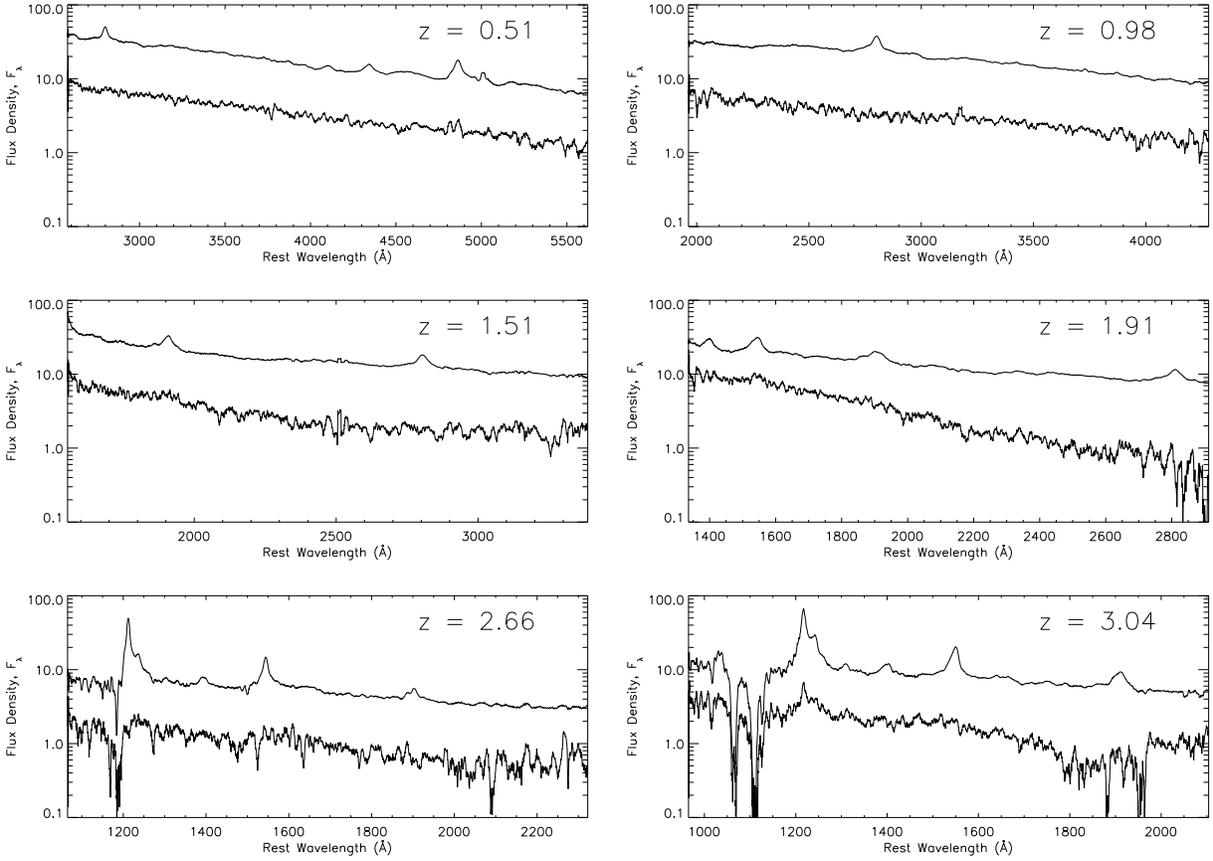}
\caption{Spectra for six representative objects from the variable quasar sample.  In each panel, the upper curve is the bright phase spectrum while the lower curve is the difference spectrum.  Redshift ($z$) is displayed for each quasar.  All flux densities are in $10^{-17}$ erg/s/cm$^{2}$/\AA.  All spectra are boxcar smoothed with a smoothing length of 20 pixels.
\label{Fig2.7}}
\end{figure}

Fig. \ref{Fig2.7} shows example spectra from a several quasars in the variable sample, ranging in redshift 
from one-half to three.  Shown are both the bright phase spectra as well as the difference spectra, defined as the 
bright phase minus the faint phase spectra.




\section{Composite Variability Spectra\label{composite}}
\subsection{Creation of Composite Spectra\label{createcomposite}}
From the objects in the variable quasar sample, we create ensemble difference 
spectra in order to study the detailed dependence of variability on wavelength.
The algorithm for creating the composite is similar to that used in \citet{vandenberk01}
to create composite quasar spectra.
  
In order to create ensemble difference spectra, we first calculate an individual 
difference spectrum for each quasar by simply subtracting the faint phase spectrum 
from the bright phase spectrum: $f_{diff}(\lambda)=f_{bright}(\lambda)-f_{faint}(\lambda)$.  
The individual difference spectra are 
shifted to the rest frame and fit by a power law ($f_{diff}(\lambda)=(\frac{\lambda}{\lambda_{0}})^{\alpha}$).
The spectra are then scaled using the power law fits such that they all have a flux density of 1 
(in arbitrary units) at a rest wavelength of 3060\AA.  (3060\AA\ is chosen as it lies in a region of the spectrum 
with little flux from line emission and is available to the SDSS spectrographs at virtually 
all redshifts).
As SDSS spectra are logarithmic in wavelength, deredshifting the spectra is tantamount 
to sliding each spectrum blueward by the appropriate number of pixels.  
Thus, rebinning the flux density is unnecessary; deredshifted spectra will be off by at most one pixel.  
Given that we cannot measure quasar redshifts to better than one pixel, this should be acceptable for creation of composite quasar spectra.
Flagged pixels are removed as described in \S\,\ref{spectro} and the arithmetic and 
geometric means are calculated for each pixel, yielding two composite difference spectra.  
Neither mean is weighted (by errors or S/N),
to avoid biasing the difference spectra towards either the most variable quasars
or those few with the highest signal-to-noise ratio spectra.  The resulting mean difference
spectra, seen in Fig. \ref{Fig3.1}, represent the average difference between 
bright- and faint-phase spectra for our variable quasar sample.

For reference, arithmetic and geometric mean composite quasar spectra are created from the high-S/N epoch 
spectra all of quasars in the variable sample, using the same 
algorithm as for the difference composite.  These composites are seen in Fig. \ref{Fig3.2}.
Bright- and faint-phase arithmetic mean composite quasar spectra (see Fig. \ref{Fig3.3}) are also created, using the same algorithm.

As discussed in \citet{vandenberk01}, arithmetic and geometric mean composites have different properties. 
The geometric mean preserves the average power law slope, insofar as quasar spectra can be accurately described by power laws.  
The arithmetic mean retains the relative strength 
of the non-power-law features, such as emission lines.  The arithmetic and geometric mean 
composites are quite similar and (except in the case of spectral slope) there is nothing to be learned from the 
geometric mean spectra that cannot be determined from their arithmetic counterparts.



\subsection{Continuum Variability\label{continuum}}




We use the difference composites, as well as the quasar composites, to study the 
ensemble properties of quasar variability.
As mentioned in \S\,\ref{createcomposite}, geometric mean composite spectra are more appropriate for studying 
spectral slopes.  Redward of 1300\AA\ (to avoid the Ly$_{\alpha}$ and NV lines), and blueward of 5800\AA\ 
(to avoid the noisiest part of the spectrum at the red end), the {\it geometric} mean difference spectrum can be fit by 
a power law:

\begin{equation}
\Delta F_{\lambda} = \Biggl(\frac{\lambda}{3060\AA}\Biggr)^{-2.00}.
\end{equation}  

A power-law fit to the geometric mean single-epoch quasar spectrum produces a more shallow 
dependence: $\alpha_{\lambda} = -1.35$. Thus, the composite difference spectrum is bluer than 
the composite quasar spectrum.  (Our geometric mean composite 
quasar spectrum is redder than the geometric mean composite from \citet{vandenberk01}, for which 
$\alpha_{\lambda} = -1.56$.  As the spectra used here were not corrected for Galactic extinction (see \S\,\ref{spectro}), 
unlike those used in \citet{vandenberk01}, this is not surprising).  

Fig. \ref{Fig3.4} shows the ratio of the arithmetic mean 
composite difference spectrum to the arithmetic mean composite quasar spectrum.  Most 
intriguing is that the ratio is near 1 (meaning the two spectra have the same continuum slope) for most 
values of rest wavelength greater than 2500\AA.  This change in the ratio means that while it is 
true that the composite difference spectrum is bluer than the average quasar spectrum, the difference 
in the slopes seems to be due to light blueward of 2500\AA.  VB04 showed that there is a strong dependence of variability 
on rest wavelength, and may have hinted that the variability dependence appears to flatten around 3000\AA\ (see Figure 13 of that paper).

These findings show that quasar variability cannot be described by a change in a simple wavelength-indepedent scale factor.  
On average, quasar continuum spectral slopes are steeper (bluer) in bright phases than faint phases.  This is not surprising; 
previously detected color changes like those seen in VB04 and \citet{trevese01} hinted at a 
change in power law slope.  The change in spectral shape appears to become stronger at wavelengths less than about 2500\AA.

\subsection{Color Variability\label{color}}

The wavelength dependence of quasar continuum variability implies a color dependence.  However, photometric color changes in variable
quasars have not always been detected.  \citet{cristiani97} found that quasars were more variable in the $B$ band than in $R$.  Similar findings were reported by \citet{hawkins03}, but the differences were attributed at least partially to a lack of variability in the spectrum of the host galaxy.  We show here that photometric color changes depend on the combined effects of continuum changes,
emission-line changes, redshift, and the selection of photometric bandpasses.

To explore the change in quasar color due to variability, we created arithmetic mean composite spectra of the 315
variable quasars in their bright and faint phases (see the upper panel Fig. \ref{Fig3.3}).  
All of the input spectra, bright and faint, were normalized to unity at 3060\AA\ during the construction of the composite spectra.  The normalization
simply removes a constant offset from the relative color changes between the bright and faint spectra.
To measure the color differences, we convolved the bright- and faint-phase 
composites with SDSS filter transmission curves at an airmass of 1.2.  Color differences 
(bright phase color minus faint-phase color) are shown as a function of redshift in Fig. \ref{Fig3.5}.  The 
bright phase composite spectrum is generally bluer than the faint-phase composite (indicated by $\Delta${color} $< 0$), 
but not always.  

The use of spectroscopy may help resolve disputes concerning color variability.  
It is clear from Fig. \ref{Fig3.5} that the apparent change in color is a function of redshift and the filters used.
Quasars are bluer in the bright phases in all colors and at most redshifts.  
However, as emission lines are redshifted in and out of the passbands, the bright phase may appear redder than 
the faint phase.  This is due to the relative lack of variability seen in the emission lines.  
In the faint phase, the emission lines are stronger, relative to the continuum, than in the 
bright phase.  When a strong emission line (such as Mg~\textsc{ii}) is redshifted into the $u$ band, for example, it has 
the effect of making the $u-g$ color of the faint phase bluer than one would expect based solely on the power-law 
continuum.  When calculating the color difference ($\Delta{(u-g)} = (u-g)_{bright} - (u-g)_{faint}$), a 
bluer-than-expected faint phase results in a red $\Delta{(u-g)}$.  In single-epoch spectroscopy, a similar effect is seen, as photometric colors
change as emission lines are shifted in and out of broad-band filters \citep{richards01}.

The first red ``bump'' in Fig. \ref{Fig3.5} (seen peaking at $z \sim 0.35$ in $\Delta{u-g}$) corresponds to the Mg~\textsc{ii} 
emission line moving through the blue filter.  The peak comes as Mg~\textsc{ii} is redshifted from the $u$ band 
into the $g$ band.  
The nadir of the first dip ($z \sim 0.7$ in $\Delta{u-g}$) corresponds to the C~\textsc{iii]} line entering the $u$ band.  
An inflection point seen in some panels ($z \sim 1.0$ in $\Delta{u-g}$) represents C~\textsc{iv} entering the $u$ band
as C~\textsc{iii]} enters the $g$ filter.  The final peak and dip correspond to C~\textsc{iv} and Ly$\alpha$ entering the $u$ 
and $g$ filters, respectively.  
The modest changes ($|\Delta${color}$| < -0.1$) seen here are consistent with the color changes 
measured in VB04, where the average quasar was $\sim0.03$ magnitudes bluer in its bright phase.

\subsection{Emission Line Variability\label{emline}}

The difference spectrum in Fig. \ref{Fig3.1} shows very little evidence of emission lines.  
"Residual" emission lines appear for C~\textsc{iv}, C~\textsc{iii]} and Mg~\textsc{ii}, but features corresponding to other lines are not clearly present.
By measuring the equivalent width of the detectable lines in the difference spectrum and comparing them to the 
equivalent widths of the same lines in the average quasar spectrum, we can determine the relative 
variability of the lines, with respect to the continuum.  We measure the equivalent widths by fitting 
straight lines to the continuum over the
same wavelength ranges used by \citet{vandenberk01} to fit emission lines in composite quasar spectra.  
The difference spectrum equivalent widths of C~\textsc{iv}, C~\textsc{iii]} and Mg~\textsc{ii} are 8.0\AA, 3.3\AA\ and 11.1\AA, 
respectively.  The equivalent widths for these lines in the average quasar spectrum are 26.9\AA, 17.7\AA\ and 36.5 \AA.  
These values indicate that the C~\textsc{iv}, C~\textsc{iii]} and Mg~\textsc{ii} lines vary only 20--30\% as much the continuum.  

The relative lack of emission-line variability is also demonstrated in Fig. \ref{Fig3.3}.
Virtually all emission lines in the upper panel of Fig. \ref{Fig3.2} show corresponding dips in variability 
in Fig. \ref{Fig3.3}.  Closer inspection of these line variability profiles shows substructure, 
including potential assymetries in the Ly$_{\alpha}$, C~\textsc{iv}, C~\textsc{iii]} and Mg~\textsc{ii} lines.  The character of the emission line 
variability is the subject of future work \citep{wilhite05}.  For now, we simply note that the broad lines are clearly less variable than the 
underlying continuum---a relationship known as the intrinsic Baldwin effect \citep[IBE;][]{kinney90}.
More discussion on the IBE's effect on color variability can be found in \S\,\ref{discussion}.




%
%

\section{Discussion\label{discussion}}

\begin{figure}
\plotone{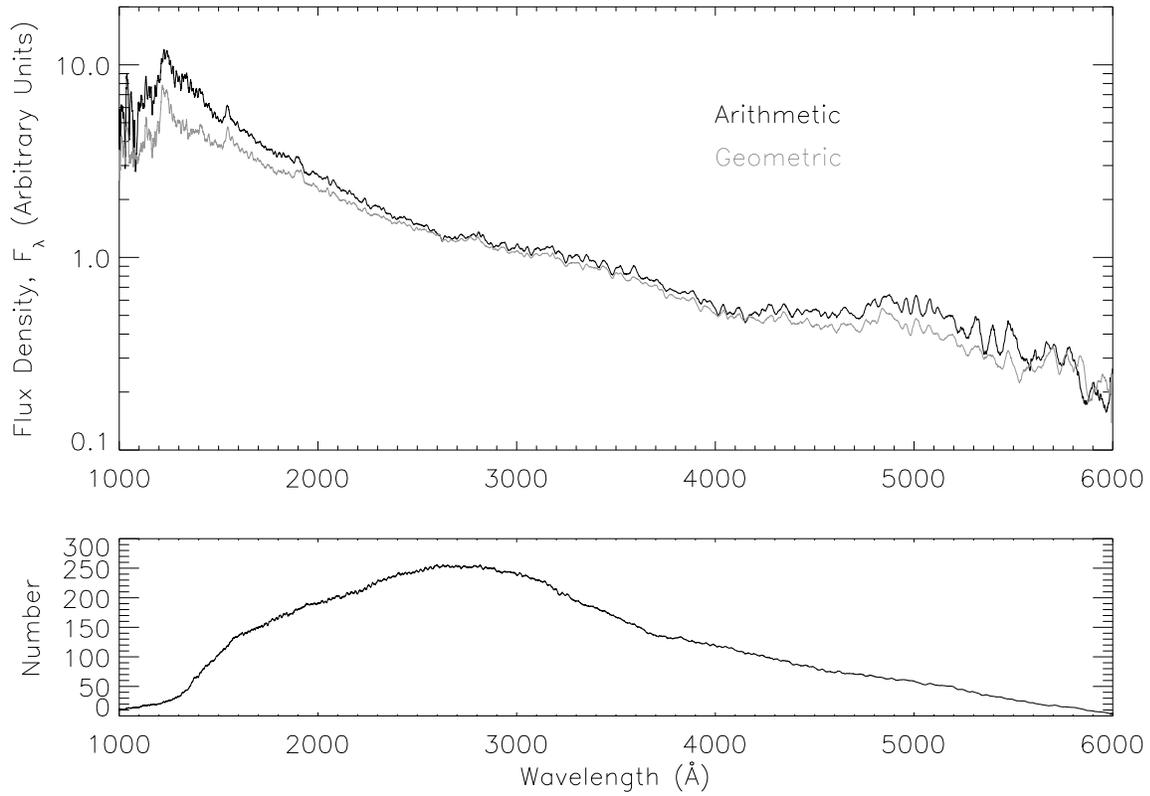}
\caption{(Upper panel) Composite difference spectra for the 315 objects in the final variable quasar sample.  The composites are created by taking the arithmetic (dark spectrum) or geometric (light) mean of the scaled flux differences.  Both spectra are boxcar smoothed with a smoothing length of 20 pixels.
(Lower Panel) Number of quasars used to create composites as a function of wavelength.  \label{Fig3.1}}
\end{figure}

\begin{figure}
\plotone{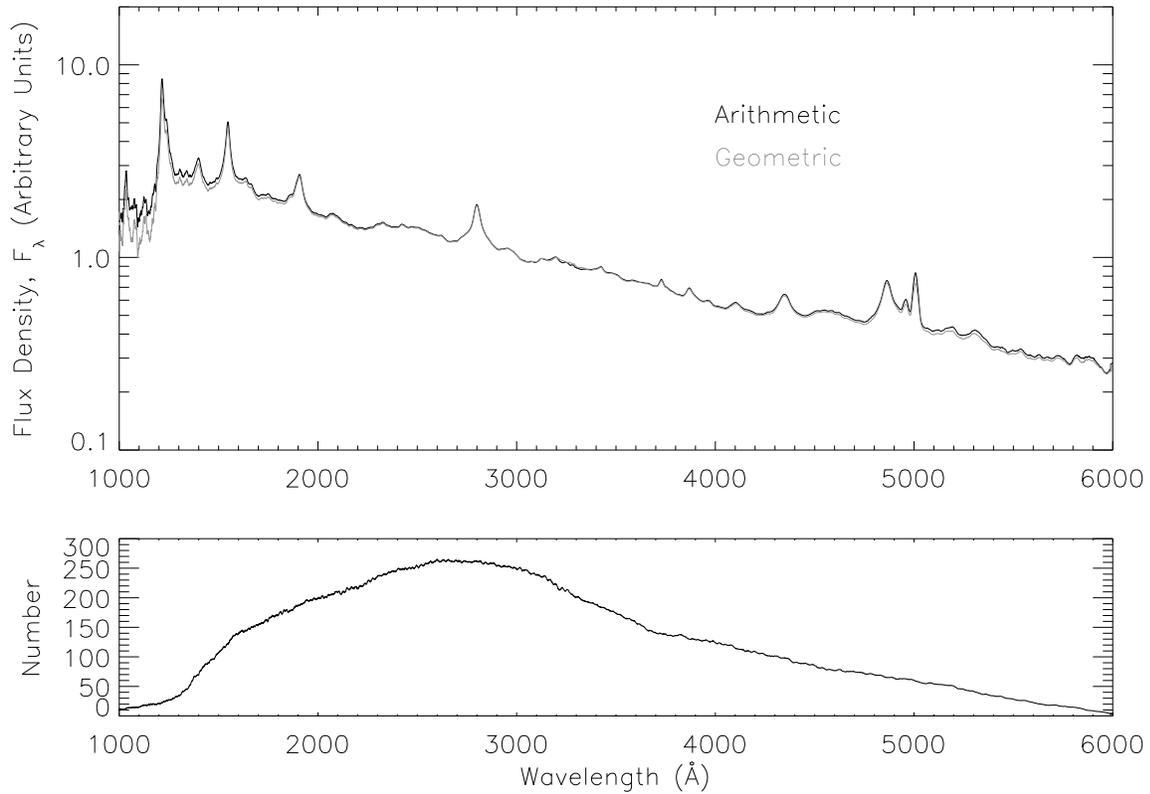}
\caption{(Upper panel) Composite single-epoch spectra for the 315 objects in the final variable quasar sample.  The composites are created by taking the arithmetic (dark spectrum) or geometric (light) mean of the scaled flux densities at the high-S/N epoch.  Both spectra are boxcar smoothed with a smoothing length of 20 pixels.  (Lower Panel) Number of quasars used to create composites as a function of wavelength.  
\label{Fig3.2}}
\end{figure}

\begin{figure}
\plotone{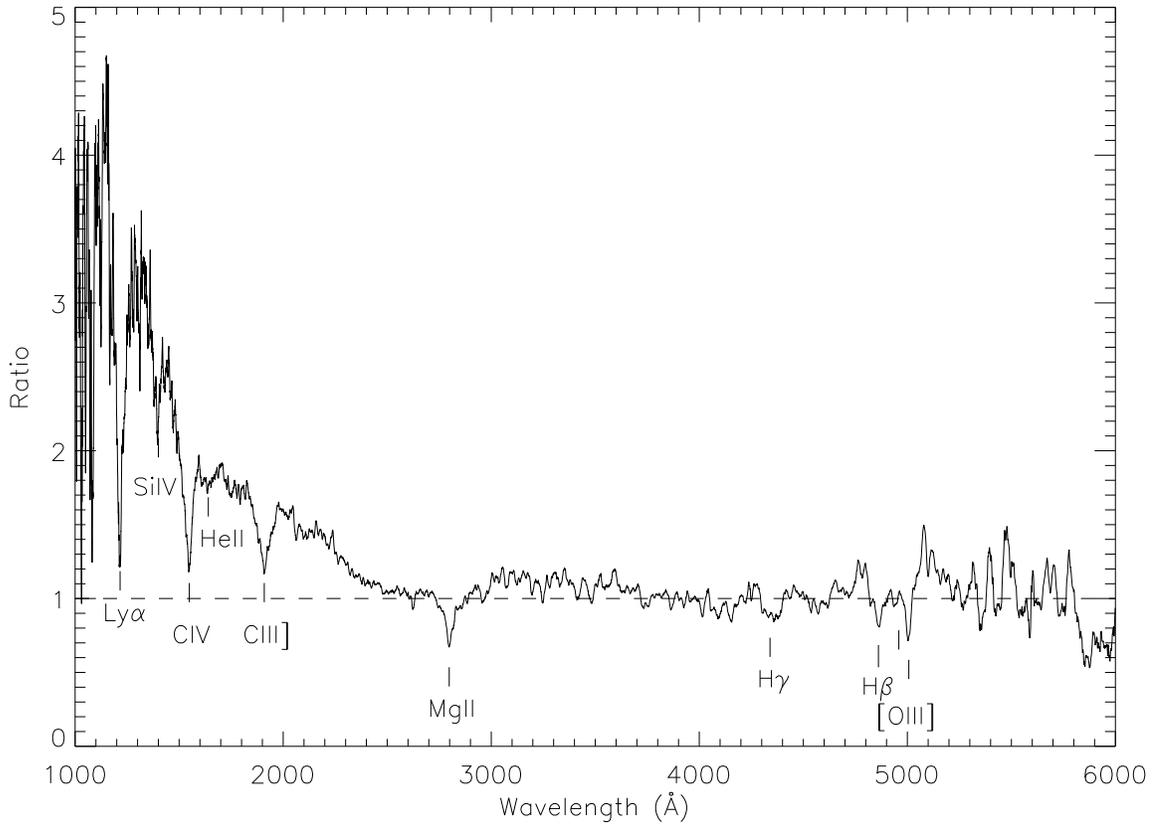}
\caption{The ratio of the arithmetic mean difference spectrum (upper panel of Fig. \ref{Fig3.1}) to the arithmetic mean singl-epoch spectrum (upper panel of Fig. \ref{Fig3.2}).  Larger values of this ratio indicate a more variable portion of the spectrum.  There appears to be a break near 2500\AA, below which the variability increases. The dips at the locations of the emission lines indicate that the lines are considerably less variable than the continuum.  
\label{Fig3.3}}
\end{figure}

\begin{figure}
\plotone{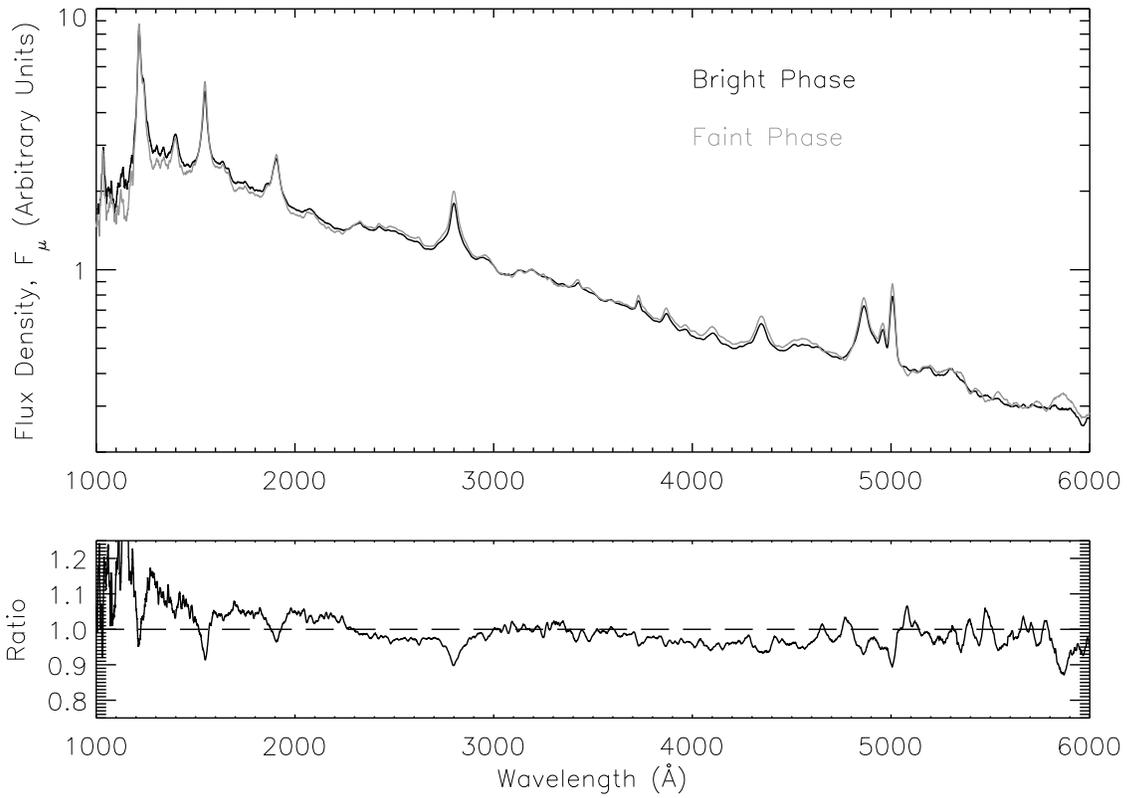}
\caption{(Upper Panel) Bright- (dark spectrum) and faint-phase (light) arithmetic mean composite spectra for the 315 objects in the final variable quasar sample.  As both composites were created from quasar spectra scaled to a value of 1 at 3060\AA, only relative color changes are meaningful.  Both spectra are boxcar smoothed with a smoothing length of 20 pixels.  (Lower panel) Ratio of bright-phase composite to faint-phase composite as a function of wavelength.  
\label{Fig3.4}}
\end{figure}

\begin{figure}
\plotone{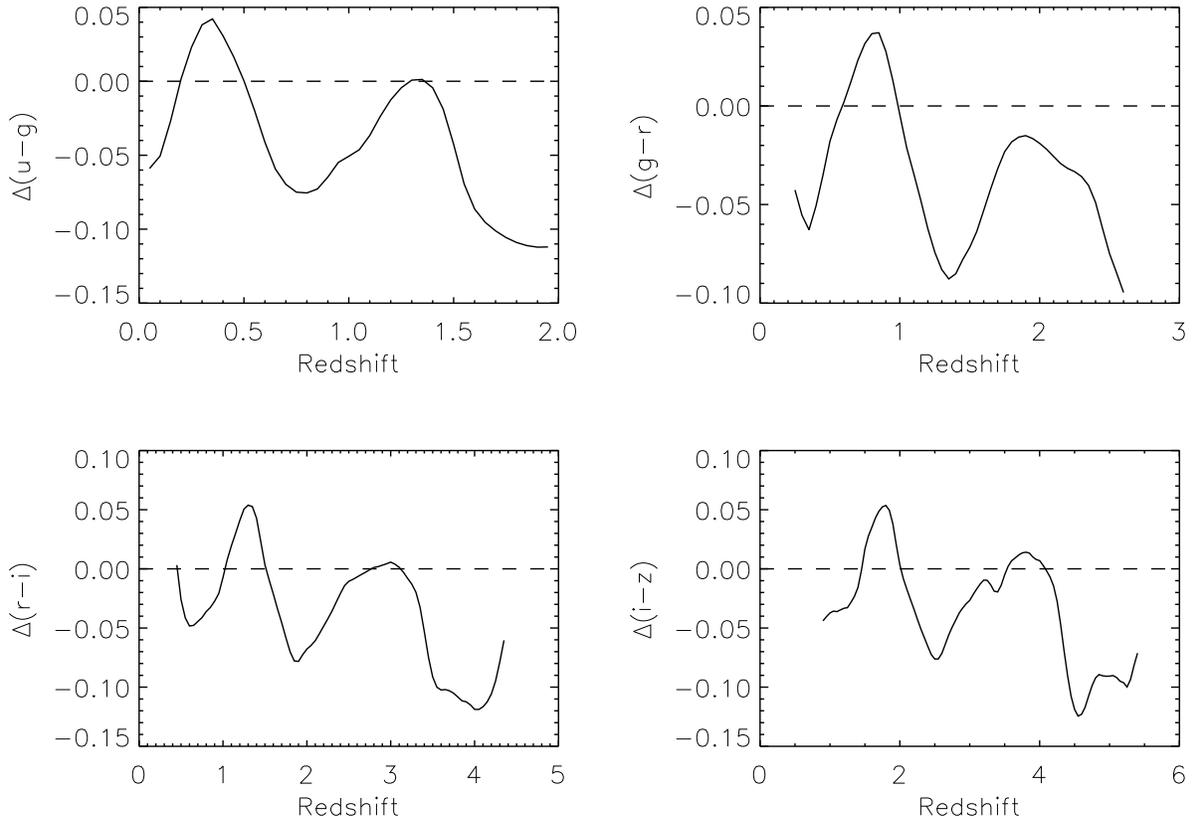}
\caption{Color difference between bright- and faint-phase composite spectra as a function of redshift.  Bright-phase quasars are bluer at most redshifts in all colors.  Peaks and valleys are due to emission lines being redshifted into and out of filter transmission curves.
\label{Fig3.5}}
\end{figure}

Although more time-consuming and hence costly, spectroscopy is superior to photometry for studying quasar 
variability in many ways.  Even the largest photometric variability studies can only produce modest 
wavelength resolution; VB04 were able to construct $\sim 100$ bins in wavelength using photometric data, while the difference spectrum in 
Fig. \ref{Fig3.1} has nearly 8000.  Additionally, spectroscopic studies allow for a probe of the variability 
of emission lines and their effects on color changes.

The difference spectrum seen in Fig. \ref{Fig3.1} is not easily fit by a single-temperature blackbody spectrum.  
High temperatures ($T \gtrsim 25,000K$) yield curves that are too steep, while lower temperature blackbody curves 
have peaks in the optical that this difference spectrum clearly does not demonstrate.  An attempt to fit a 
single-temperature blackbody curve yields a $\chi^{2}$ value 50 times greater than that for a single power-law 
fit.  This implies that a single-temperature "hot spot" is not a good model for variability.  An upcoming paper 
\citep{pereyra04} will examine whether these continuum variations can be explained by standard accretion disk models \citep[e.g.,][]{shakura73}.
Continuum luminosity and slope changes are also important for studies of the broad line region.  A change in the ionizing spectrum shape may be necessary to explain some broad-line properties \citep{korista04}.

In principle, a number of physical mechanisms could be responsible for the wavelength dependence of 
variability seen here.  Poissonian models predict that quasar variability (as well as the bulk of quasar 
luminosity) is due to random, unassociated events, such as supernova explosions \citep[e.g.,][]{terlevich92,torricelli00}.  
They do predict a bluer spectra for bright-phase quasars, but this color change is due entirely to the change in relative importance 
of the underlying host galaxy spectrum as the quasar brightens and fades \citep{cid00}.  Given that the median 
absolute $r$-band magnitude of quasars in our variable sample is $\sim-25$, some 30 times brighter than an 
$M_{r,*}$ galaxy \citep{blanton03b}, it is unlikely that the galaxy light could be responsible for the strong 
wavelength dependence seen in Fig. \ref{Fig3.1}.  
Microlensing models can qualitatively account for bluer spectra
and weaker emission line variability.  If the BLR is large relative to
the Einstein radius of the lens, it will not be magnified as much as a more compact
region (unless there is significant substructure within the BLR).  
If the continuum source is compact, and is bluer at smaller
radii, the bluer parts may be magnified more than the redder (larger)
parts.  However, one would also expect more numerous cases in
which the lens only crosses the extended BLR, and not the compact continuum source,
and only the emission lines are magnified.  The sample used here would not contain those objects which had shown line variability without variability in the continuum, but this could be tested with the full sample of $\sim$ 2000 multiply-observed quasars.

Reverberation mapping of high-luminosity quasars suffers from two main difficulties, both related to the increased size of the broad-line region in these objects.  The BLR radius scales with continuum luminosity with an exponent ranging from 0.5 \citep{wandel99} to 0.7 \citep{kaspi00}.  First, a larger BLR leads to less coherent variability in the emission lines.  The size of the BLR is presumed to be directly related to the range of light travel times from the central source to the BLR and on to the observer.  Thus, the continuum fluctuations which drive the emission line variability are "smeared out" in time.  Secondly, the BLR response time is directly proportional to the radius of the BLR.  Thus, not only are the emission line fluctuations small in a high-luminosity quasar, observers must wait longer before they are evident.
The composite difference spectrum in Fig. \ref{Fig3.1} offers both hope and caution for future attempts at reverberation mapping of high-luminosity quasars.
Since a few of the lines (C~\textsc{iv}, C~\textsc{iii]}, Mg~\textsc{ii}) are clearly observed in this spectrum, there is hope that the emission line variations needed for reverberation mapping are detectable, though small.  Additionally, this sample of 315 highly variable quasars offers a large list of candidates for future high-luminosity reverberation-mapping campaigns.

\section{Conclusions\label{conclusions}}

We have explored the detailed wavelength dependence of quasar variability.  
This marks the first time that the wavelength dependence has been studied at 
high enough resolution to resolve the ensemble variability properties of 
quasar emission lines.  Using stars as an assumed non-variable population, we have recalibrated the SDSS
spectrophotometry for all quasars on 53 multiply observed spectroscopic plates.  By comparison with the 
assumed non-variable portion of the stellar population, we have 
selected those quasars which have varied significantly in total flux between observations.  
From composite difference spectra created from these 315 quasars, we can draw four conclusions about 
ultraviolet/optical quasar variability:

1) A comparison of the spectral slopes reveals that the composite difference spectrum is 
bluer than the composite quasar spectrum.  This change is much stronger at wavelengths shorter than $\lambda_{rest} = 2500$\AA.

2) The flux in the emission lines is significantly less variable than the continuum.  This is demonstrated by both the 
composite difference spectrum and the ratio of the difference spectrum to the average quasar spectrum.  
It is also seen in the difference spectra of individual objects.  The relative lack of emission line variability with
respect to the continuum results in the Intrinsic Baldwin Effect.

3) While quasars are generally bluer in their bright phase, the specific color change seen in 
photometry is a function of quasar redshift and the filters used.  The relative lack of variability of the 
emission lines is such that, for some ranges in redshift, bright-phase quasars may appear redder in 
photometric measurements than when in the faint phase.

4) The shape of the difference spectrum is inconsistent with a single-temperature blackbody, but is well described by a power law.
The slope of the difference spectrum is significantly steeper (bluer) than that of the average quasar, indicating that the bright-phase continuum
is not simply a scaled version of the faint-phase continuum. 

Many possibilities exist for future variability studies using this dataset.  The number of objects 
offers a large dynamic range in other properties (such as luminosity or redshift), which can be used as 
levers to study the detailed wavelength dependence of variability as a function of these properties.  
Another intriguing possibility is that difference spectra, which are virtually devoid of emission lines and 
non-variable spectral components (such as the underlying galaxy spectrum) may offer a ``clean'' means by which 
to study the quasar continuum.
Multiple-epoch spectra may also yield information on the structure of the broad line region or 
the validity of emission-line orientation measures.  Additionally, this set of 315 variable quasars 
represents a potentially useful target list for future use in high-redshift, high-luminsoty reverberation mapping studies.

D.E.V. and D.P.S. wish to acknowledge the support of NSF grant AST03-07582.

Funding for the creation and distribution of the SDSS Archive has been provided by the Alfred P. Sloan Foundation, the 
Participating Institutions, the National Aeronautics and Space Administration, the National Science Foundation, the U.S. 
Department of Energy, the Japanese Monbukagakusho, and the Max Planck Society. The SDSS Web site is http://www.sdss.org/.

The SDSS is managed by the Astrophysical Research Consortium (ARC) for the Participating Institutions. The Participating Institutions are The University of Chicago, Fermilab, the Institute for Advanced Study, the Japan Participation Group, The Johns Hopkins University, the Korean Scientist Group, Los Alamos National Laboratory, the Max-Planck-Institute for Astronomy (MPIA), the Max-Planck-Institute for Astrophysics (MPA), New Mexico State University, University of Pittsburgh, Princeton University, the United States Naval Observatory, and the University of Washington.


\onecolumn
\clearpage
%



\clearpage

\clearpage

\clearpage

\clearpage

\clearpage

\clearpage

\clearpage

\clearpage


\clearpage

\clearpage

\clearpage

\clearpage

\clearpage

\clearpage



\begin{deluxetable}{cccccc} 
\tablewidth{0pt}
\tablecaption{SDSS plates with multiple observations separated by more than 50 days.  Asterisks indicate that a data release includes observations from an MJD not used in this paper.
\label{platetable}}

\tablehead{
	\colhead{} &
	\multicolumn{2}{c}{High-S/N Epoch} & 
	\multicolumn{2}{c}{Low-S/N Epoch} & 
	\colhead{$\Delta{\tau}$}\\
	\colhead{Plate} &
	\colhead{MJD} & 
	\colhead{Data Release} & 
	\colhead{MJD} & 
	\colhead{Data Release} & 
	\colhead{(days)}
}
\tablecolumns{6}
\startdata

269  &  51910  &  DR1  &  51581  &       &  329  \\  
279  &  51984  &  DR1  &  51608  &       &  376  \\  
283  &  51959  &  DR1  &  51584  &       &  375  \\  
284  &  51943  &  DR1  &  51662  &       &  281  \\  
285  &  51930  &  DR1  &  51663  &       &  267  \\  
291  &  51928  &  DR1  &  51660  &       &  268  \\  
293  &  51689  &  DR1  &  51994  &  DR2  &  305  \\  
295  &  51985  &  DR1  &  51585  &       &  400  \\  
296  &  51984  &  DR1  &  51665  &       &  319  \\  
297  &  51959  &  DR1  &  51663  &       &  296  \\  
298  &  51955  &  DR1  &  51662  &       &  293  \\  
300  &  51943  &  DR1  &  51666  &       &  277  \\  
301  &  51942  &  DR1  &  51641  &       &  301  \\  
304  &  51609  &       &  51957  &  DR1  &  348  \\  
306  &  51637  &  DR1  &  51690  &       &   53  \\  
309  &  51994  &  DR1  &  51666  &       &  328  \\  
310  &  51990  &  DR1  &  51614  &       &  376  \\  
340  &  51990  &  DR1  &  51691  &       &  299  \\  
351  &  51695  &       &  51780  &  DR1  &   85  \\  
352  &  51694  &  DR1  &  51789  &  DR2  &   95  \\  
385  &  51977  &  DR1  &  51783  &       &  94   \\  
390  &  51900  &  DR1  &  51816  &       &  84   \\  
394  &  51876  &  DR1* &  51812  &       &  101  \\  
404  &  51812  &  DR1  &  51877  &       &  65   \\  
406  &  51817  &  DR1* &  52238  &       &  421  \\  
410  &  51816  &  DR1  &  51877  &       &  61   \\  
411  &  51817  &  DR1  &  51914  &       &  97   \\  
412  &  51931  &  DR1  &  52258  &  DR2  &  327  \\  
413  &  51929  &  DR1  &  51821  &       &  108  \\  
415  &  51810  &  DR1  &  51879  &       &  69   \\  
416  &  51811  &  DR1  &  51885  &       &  74   \\  
418  &  51817  &  DR1  &  51884  &       &  67   \\  
419  &  51879  &  DR1  &  51812  &       &  67   \\  
422  &  51811  &  DR1  &  51878  &       &  67   \\  
476  &  52314  &  DR2  &  52027  &       &  287  \\  
525  &  52295  &  DR2  &  52029  &       &  266  \\  
547  &  52207  &  DR2  &  51959  &       &  248  \\  
616  &  52374  &  DR2  &  52442  &       &  68   \\  
620  &  52375  &  DR2  &  52081  &       &  294  \\  
678  &  52884  &       &  52534  &       &  350  \\  
739  &  52520  &  DR3  &  52264  &       &  256  \\  
756  &  52237  &  DR3  &  52577  &       &  340  \\  
790  &  52346  &  DR2* &  52433  &       &  95   \\  
791  &  52347  &       &  52435  &  DR2  &  88   \\  
803  &  52318  &       &  52264  &       &  54   \\  
810  &  52672  &       &  52326  &       &  346  \\  
814  &  52443  &  DR2  &  52355  &       &  88   \\  
873  &  52347  &       &  52674  &  DR3  &  327  \\  
876  &  52346  &       &  52669  &  DR3  &  323  \\  
889  &  52346  &       &  52663  &  DR3  &  317  \\  
1034  &  52813  &       &  52525  &       &  288  \\  
1037  &  52878  &       &  52826  &       &  52   \\  
1296  &  52962  &       &  52737  &       &  225  \\ 
\enddata
\end{deluxetable}


\begin{deluxetable}{lcccrccrr}
\tablewidth{0pt}
\tablecaption{Variabile Quasar Sample.  HSN and LSN indicate the high- and low-S/N ratio epochs, respectively.  All redshifts and magnitudes are from the high-S/N ratio epoch. 
\label{qsotable}}
\tablehead{
	\colhead{SDSS J} & 
	\multicolumn{2}{c}{MJD} & 
	\colhead{$z_{HSN}$} & 
	\colhead{$\Delta{\tau}$} & 
  \colhead{$r_{HSN}$} &
	\colhead{$M_{r,HSN}$} & 
	\multicolumn{2}{c}{$S/N_{r}$}\\
	\colhead{} &
	\colhead{HSN} &
	\colhead{LSN} &
	\colhead{} &
	\colhead{(days)} &
	\colhead{} &
	\colhead{} &
	\colhead{HSN} &
	\colhead{LSN}
}
\tablecolumns{9}
\startdata

       100013.37+011203.2 & 51910 & 51581 &  1.80 & 117.4 &  19.1 &  $-26.0$ &  12.3 &  11.4 \\

       100428.43+001825.6 & 51910 & 51581 &  3.04 &  81.3 &  18.7 &  $-27.6$ &  16.4 &  11.2 \\

       100517.01+005905.9 & 51910 & 51581 &  0.69 & 194.3 &  20.1 &  $-22.7$ &   6.1 &   2.7 \\

       100514.76+005341.2 & 51910 & 51581 &  0.95 & 169.1 &  19.4 &  $-24.2$ &  10.8 &   6.8 \\

     111506.78$-$002255.3 & 51984 & 51608 &  1.35 & 160.1 &  18.3 &  $-26.1$ &  24.0 &  22.7 \\

     111400.00$-$002342.3 & 51984 & 51608 &  0.96 & 192.2 &  19.7 &  $-23.8$ &   8.5 &   7.3 \\

       111221.82+003028.5 & 51984 & 51608 &  0.52 & 246.8 &  19.6 &  $-22.5$ &  11.3 &   4.6 \\

     114538.17$-$010010.5 & 51959 & 51584 &  0.88 & 199.4 &  19.5 &  $-23.9$ &  11.1 &   8.9 \\

     114528.56$-$004739.1 & 51959 & 51584 &  0.72 & 218.6 &  19.4 &  $-23.5$ &  12.7 &   9.9 \\

     114211.59$-$005344.2 & 51959 & 51584 &  1.92 & 128.5 &  18.4 &  $-26.8$ &  22.7 &  16.4 \\

     114117.63$-$001250.4 & 51959 & 51584 &  0.50 & 249.3 &  18.1 &  $-23.9$ &  26.4 &  12.1 \\

       114016.67+005351.3 & 51959 & 51584 &  1.13 & 175.7 &  18.1 &  $-25.9$ &  25.8 &  15.9 \\

       114354.02+011343.1 & 51959 & 51584 &  1.28 & 164.8 &  17.5 &  $-26.7$ &  31.2 &  23.8 \\

     114533.85$-$000933.0 & 51959 & 51584 &  1.04 & 183.5 &  18.9 &  $-24.9$ &  18.0 &   9.6 \\

       114612.09+002105.1 & 51959 & 51584 &  1.23 & 168.4 &  18.9 &  $-25.2$ &  18.7 &   8.5 \\

       114510.38+011056.2 & 51959 & 51584 &  0.63 & 230.6 &  20.3 &  $-22.3$ &   5.6 &   4.2 \\

     114749.66$-$000109.4 & 51959 & 51584 &  1.26 & 166.1 &  18.8 &  $-25.4$ &  16.5 &   7.9 \\

       114948.81+000855.8 & 51959 & 51584 &  1.97 & 126.3 &  19.1 &  $-26.2$ &  12.9 &  10.9 \\

     115154.83$-$005904.6 & 51943 & 51662 &  1.93 &  96.0 &  19.2 &  $-26.0$ &  12.9 &   7.6 \\

     115137.20$-$005013.0 & 51943 & 51662 &  1.49 & 112.8 &  18.8 &  $-25.8$ &  18.4 &  11.7 \\

     115240.52$-$003004.3 & 51943 & 51662 &  0.55 & 181.0 &  19.4 &  $-22.8$ &  10.9 &  10.6 \\

     114957.56$-$010634.5 & 51943 & 51662 &  1.14 & 131.1 &  19.4 &  $-24.6$ &  12.0 &   8.1 \\

     115043.87$-$002354.0 & 51943 & 51662 &  1.98 &  94.4 &  17.5 &  $-27.8$ &  40.0 &  27.2 \\

       114931.53+004242.2 & 51943 & 51662 &  0.93 & 145.5 &  19.1 &  $-24.4$ &  15.9 &   7.6 \\

       115057.91+005718.1 & 51943 & 51662 &  1.27 & 123.5 &  19.3 &  $-25.0$ &  12.1 &   6.8 \\

       115213.55+001946.7 & 51943 & 51662 &  1.83 &  99.2 &  19.8 &  $-25.3$ &   8.0 &   7.3 \\

       115253.68+000131.8 & 51943 & 51662 &  0.82 & 154.2 &  19.5 &  $-23.7$ &  10.0 &   8.9 \\

     115412.54$-$000153.2 & 51943 & 51662 &  0.73 & 162.4 &  17.8 &  $-25.1$ &  32.6 &  26.7 \\

     115403.28$-$005253.4 & 51930 & 51663 &  0.75 & 152.8 &  19.5 &  $-23.5$ &  12.5 &   8.1 \\

       120123.24+002828.3 & 51930 & 51663 &  1.37 & 112.7 &  18.8 &  $-25.6$ &  21.4 &  11.6 \\

       120142.98+004924.8 & 51930 & 51663 &  1.52 & 105.9 &  18.4 &  $-26.2$ &  27.2 &  15.3 \\

     124555.11$-$003735.3 & 51928 & 51660 &  1.04 & 131.2 &  18.4 &  $-25.4$ &  25.8 &  15.6 \\

     124524.59$-$000937.9 & 51928 & 51660 &  2.08 &  86.9 &  18.2 &  $-27.3$ &  29.7 &  20.8 \\

     124356.22$-$000021.8 & 51928 & 51660 &  1.84 &  94.5 &  19.6 &  $-25.5$ &   9.8 &   7.2 \\

       124242.11+001157.9 & 51928 & 51660 &  2.16 &  84.8 &  19.2 &  $-26.3$ &  14.2 &   9.4 \\

     130258.48$-$002603.4 & 51689 & 51994 &  1.00 & 152.3 &  19.9 &  $-23.7$ &   5.7 &   5.6 \\

     125737.06$-$003220.1 & 51689 & 51994 &  1.03 & 150.5 &  19.2 &  $-24.6$ &  14.7 &  11.1 \\

     125658.39$-$002122.9 & 51689 & 51994 &  1.27 & 134.1 &  18.8 &  $-25.4$ &  18.3 &  12.6 \\

     125550.30$-$001831.0 & 51689 & 51994 &  0.72 & 177.6 &  18.3 &  $-24.6$ &  24.9 &  24.8 \\

     125531.95$-$002850.4 & 51689 & 51994 &  1.49 & 122.7 &  19.4 &  $-25.2$ &  11.8 &   9.8 \\

       125607.85+000207.7 & 51689 & 51994 &  0.65 & 184.8 &  18.6 &  $-24.1$ &  21.9 &  17.1 \\

     125617.52$-$001918.2 & 51689 & 51994 &  1.77 & 110.1 &  19.9 &  $-25.1$ &   7.9 &   5.5 \\

     125532.24$-$010608.7 & 51689 & 51994 &  1.78 & 109.5 &  19.8 &  $-25.2$ &   8.5 &  10.3 \\

     125209.66$-$001553.4 & 51689 & 51994 &  0.81 & 168.2 &  18.7 &  $-24.5$ &  19.4 &  13.3 \\

     125258.75$-$004236.0 & 51689 & 51994 &  1.37 & 128.6 &  19.7 &  $-24.8$ &   9.6 &   9.9 \\

       125414.43+001946.5 & 51689 & 51994 &  1.54 & 120.1 &  19.8 &  $-24.9$ &   7.3 &   9.8 \\

       125703.12+002435.9 & 51689 & 51994 &  1.26 & 135.0 &  17.6 &  $-26.6$ &  35.7 &  30.2 \\

       125629.48+002830.1 & 51689 & 51994 &  0.51 & 201.7 &  18.2 &  $-23.8$ &  24.9 &  20.6 \\

     131744.92$-$001250.4 & 51985 & 51585 &  0.92 & 208.7 &  18.2 &  $-25.2$ &  22.9 &  13.6 \\

     131433.49$-$005017.4 & 51985 & 51585 &  0.63 & 245.0 &  20.3 &  $-22.2$ &   4.5 &   3.1 \\

     131131.02$-$010332.3 & 51985 & 51585 &  1.30 & 173.6 &  19.0 &  $-25.3$ &  13.3 &  14.4 \\

     131315.18$-$000624.2 & 51985 & 51585 &  1.19 & 182.3 &  19.3 &  $-24.8$ &  11.5 &   8.7 \\

       131552.10+003803.2 & 51985 & 51585 &  1.29 & 174.5 &  19.1 &  $-25.2$ &  10.8 &   9.7 \\

       131452.78+011120.3 & 51985 & 51585 &  1.36 & 169.2 &  19.0 &  $-25.5$ &  14.9 &   8.0 \\

       131558.97+005511.2 & 51985 & 51585 &  1.57 & 155.5 &  19.4 &  $-25.4$ &  11.8 &  11.1 \\

       131630.46+005125.5 & 51985 & 51585 &  2.40 & 117.5 &  18.6 &  $-27.2$ &  18.3 &  11.7 \\

     132235.99$-$004912.4 & 51984 & 51665 &  1.14 & 149.3 &  18.2 &  $-25.8$ &  23.8 &  19.4 \\

       131840.95+003103.9 & 51984 & 51665 &  1.77 & 115.0 &  19.8 &  $-25.3$ &   8.2 &   6.8 \\

       132110.81+003821.6 & 51984 & 51665 &  4.72 &  55.8 &  21.6 &  $-25.6$ &   1.7 &   0.9 \\

       132206.20+001759.7 & 51984 & 51665 &  0.53 & 208.5 &  19.0 &  $-23.1$ &  13.0 &  13.8 \\

       132251.65+004654.8 & 51984 & 51665 &  0.52 & 209.8 &  18.8 &  $-23.2$ &  17.4 &  10.3 \\

     132832.50$-$010318.1 & 51959 & 51663 &  0.74 & 170.4 &  19.3 &  $-23.6$ &  11.2 &   7.2 \\

     132946.25$-$001805.5 & 51959 & 51663 &  1.02 & 146.4 &  19.7 &  $-24.0$ &   8.6 &   5.2 \\

       132228.48+000235.4 & 51959 & 51663 &  1.60 & 113.9 &  19.6 &  $-25.2$ &   9.8 &   8.3 \\

       132214.82+005419.9 & 51959 & 51663 &  2.15 &  94.1 &  18.6 &  $-26.8$ &  19.6 &  15.0 \\

       132717.64+005749.0 & 51959 & 51663 &  0.63 & 181.2 &  19.5 &  $-23.1$ &  11.3 &  10.2 \\

     133526.01$-$010028.1 & 51955 & 51662 &  0.67 & 175.2 &  17.8 &  $-24.9$ &  32.7 &  25.0 \\

       133127.95+000153.0 & 51955 & 51662 &  0.67 & 175.3 &  19.8 &  $-22.9$ &   6.8 &   7.5 \\

     132910.66$-$001900.5 & 51955 & 51662 &  0.56 & 188.4 &  19.4 &  $-22.9$ &  10.2 &   5.2 \\

       133044.79+011159.2 & 51955 & 51662 &  1.03 & 144.1 &  18.9 &  $-24.9$ &  14.5 &   9.2 \\

       133321.90+005824.3 & 51955 & 51662 &  1.51 & 116.6 &  18.2 &  $-26.5$ &  25.3 &  19.0 \\

       133818.56+004915.7 & 51955 & 51662 &  1.53 & 115.6 &  19.2 &  $-25.5$ &  10.2 &   7.2 \\

       133906.12+004137.9 & 51955 & 51662 &  1.22 & 132.1 &  19.5 &  $-24.6$ &   8.2 &   5.5 \\

       133939.01+001021.6 & 51955 & 51662 &  2.13 &  93.7 &  19.4 &  $-26.1$ &  10.4 &   4.6 \\

     134934.30$-$004102.8 & 51943 & 51666 &  0.51 & 182.8 &  18.2 &  $-23.9$ &  27.4 &  23.3 \\

     134855.04$-$000750.8 & 51943 & 51666 &  0.50 & 184.3 &  19.9 &  $-22.1$ &   6.5 &   5.9 \\

     134412.88$-$003005.4 & 51943 & 51666 &  0.71 & 162.2 &  20.9 &  $-22.0$ &   3.0 &   3.7 \\

     134425.95$-$000056.1 & 51943 & 51666 &  1.10 & 132.2 &  19.0 &  $-24.9$ &  13.9 &  17.0 \\

       134655.94+005700.7 & 51943 & 51666 &  1.43 & 113.9 &  19.0 &  $-25.5$ &  14.9 &  11.7 \\

       134854.42+002953.4 & 51943 & 51666 &  1.08 & 133.0 &  19.4 &  $-24.4$ &  10.7 &  13.2 \\

       135128.31+010338.6 & 51943 & 51666 &  1.09 & 132.8 &  17.3 &  $-26.6$ &  40.2 &  34.6 \\

     140248.12$-$004924.0 & 51942 & 51641 &  1.59 & 116.1 &  20.0 &  $-24.8$ &   7.7 &   7.3 \\

     140114.28$-$004537.1 & 51942 & 51641 &  2.52 &  85.5 &  19.1 &  $-26.7$ &  14.2 &  12.8 \\

     140019.73$-$004747.9 & 51942 & 51641 &  1.21 & 136.4 &  18.2 &  $-25.9$ &  26.4 &  22.3 \\

     140103.31$-$005030.6 & 51942 & 51641 &  0.93 & 156.2 &  18.8 &  $-24.7$ &  17.2 &  19.0 \\

     135844.57$-$011055.1 & 51942 & 51641 &  1.96 & 101.6 &  19.0 &  $-26.3$ &  16.7 &  18.0 \\

     135605.41$-$010024.4 & 51942 & 51641 &  1.89 & 104.3 &  19.4 &  $-25.8$ &  12.6 &   8.9 \\

     135247.96$-$002351.6 & 51942 & 51641 &  1.67 & 112.7 &  20.2 &  $-24.7$ &   6.3 &   4.5 \\

       135535.82+004213.0 & 51942 & 51641 &  1.62 & 114.9 &  18.8 &  $-26.0$ &  17.6 &  13.6 \\

       135650.33+010244.3 & 51942 & 51641 &  1.21 & 136.0 &  17.7 &  $-26.5$ &  33.0 &  27.9 \\

       135742.85+005023.9 & 51942 & 51641 &  1.58 & 116.6 &  20.6 &  $-24.1$ &   4.2 &   2.4 \\

       135533.15+002358.1 & 51942 & 51641 &  1.57 & 117.2 &  19.5 &  $-25.2$ &  10.4 &   8.3 \\

       135828.74+005811.5 & 51942 & 51641 &  3.92 &  61.1 &  19.7 &  $-27.1$ &   9.5 &   7.1 \\

       135703.87+000515.7 & 51942 & 51641 &  1.01 & 149.4 &  19.1 &  $-24.6$ &  15.2 &  17.6 \\

       135726.48+001542.3 & 51942 & 51641 &  0.66 & 181.1 &  18.8 &  $-23.9$ &  19.2 &  21.0 \\

       140112.03+003759.3 & 51942 & 51641 &  1.63 & 114.6 &  19.0 &  $-25.9$ &  17.8 &  13.6 \\

     142124.12$-$000216.5 & 51609 & 51957 &  1.16 & 161.3 &  19.4 &  $-24.6$ &  10.4 &   8.6 \\

       142053.11+001450.0 & 51609 & 51957 &  1.50 & 139.4 &  19.2 &  $-25.4$ &  12.8 &   8.9 \\

     142253.31$-$000148.9 & 51609 & 51957 &  1.08 & 167.1 &  19.4 &  $-24.5$ &  10.9 &  13.6 \\

     142205.10$-$000120.7 & 51609 & 51957 &  1.86 & 121.7 &  19.1 &  $-26.0$ &  13.4 &  11.2 \\

       142209.11+005436.3 & 51609 & 51957 &  3.68 &  74.3 &  19.7 &  $-26.9$ &   8.3 &   9.1 \\

     145555.00$-$003713.4 & 51994 & 51666 &  1.95 & 111.3 &  19.8 &  $-25.4$ &   9.3 &   5.5 \\

     145437.83$-$003706.6 & 51994 & 51666 &  0.58 & 208.2 &  19.9 &  $-22.4$ &   9.2 &   9.0 \\

     145454.69$-$000514.3 & 51994 & 51666 &  1.41 & 135.8 &  19.0 &  $-25.5$ &  18.3 &   9.0 \\

     145302.09$-$010524.4 & 51994 & 51666 &  1.81 & 116.8 &  18.2 &  $-26.9$ &  27.6 &  17.9 \\

       145246.52+003450.5 & 51994 & 51666 &  2.54 &  92.6 &  21.3 &  $-24.6$ &   2.8 &   4.0 \\

       145429.65+004121.2 & 51994 & 51666 &  2.66 &  89.6 &  19.3 &  $-26.7$ &  14.1 &   9.7 \\

     150314.57$-$000905.8 & 51990 & 51614 &  1.70 & 139.2 &  19.6 &  $-25.3$ &   9.3 &   7.8 \\

     150438.84$-$001839.4 & 51990 & 51614 &  1.16 & 173.7 &  19.0 &  $-25.0$ &  14.2 &   8.8 \\

     150150.93$-$005628.0 & 51990 & 51614 &  0.65 & 227.4 &  19.1 &  $-23.5$ &  12.3 &  10.8 \\

       145958.72+011100.3 & 51990 & 51614 &  1.24 & 167.6 &  19.5 &  $-24.7$ &   9.4 &  10.1 \\

     131728.74$-$024759.4 & 51691 & 51990 &  3.38 &  68.3 &  19.8 &  $-26.7$ &   8.5 &   9.1 \\

     131550.18$-$024357.2 & 51691 & 51990 &  0.70 & 175.4 &  18.9 &  $-24.0$ &  15.5 &  13.0 \\

     131251.79$-$024737.5 & 51691 & 51990 &  0.87 & 159.7 &  18.5 &  $-24.8$ &  20.3 &  17.8 \\

       172909.93+624519.7 & 51694 & 51789 &  1.75 &  34.6 &  19.2 &  $-25.8$ &  10.6 &   8.2 \\

       171130.49+633602.1 & 51694 & 51789 &  1.46 &  38.6 &  19.4 &  $-25.1$ &   9.5 &  11.2 \\

     234722.35$-$000921.5 & 51877 & 51783 &  1.53 &  37.2 &  21.1 &  $-23.6$ &   2.9 &   2.9 \\

     234145.50$-$004640.6 & 51877 & 51783 &  0.52 &  61.7 &  18.2 &  $-24.0$ &  30.4 &  25.1 \\

       022758.19+000225.3 & 51817 & 52238 &  1.06 & 204.0 &  20.1 &  $-23.8$ &   6.6 &   7.8 \\

     022606.35$-$005429.0 & 51817 & 52238 &  1.61 & 161.2 &  19.7 &  $-25.1$ &   8.4 &  10.7 \\

     022638.73$-$000557.9 & 51817 & 52238 &  1.08 & 202.5 &  19.3 &  $-24.5$ &  11.6 &   9.1 \\

       022534.09+000347.9 & 51817 & 52238 &  1.74 & 153.9 &  20.8 &  $-24.2$ &   4.0 &   2.8 \\

     022214.56$-$000321.7 & 51817 & 52238 &  1.07 & 203.8 &  19.3 &  $-24.5$ &  11.7 &   8.4 \\

     022222.44$-$001318.4 & 51817 & 52238 &  1.43 & 172.9 &  19.3 &  $-25.2$ &   8.8 &   7.7 \\

     022252.26$-$000942.3 & 51817 & 52238 &  0.81 & 232.1 &  20.2 &  $-23.0$ &   5.6 &   2.3 \\

     022214.38$-$001745.2 & 51817 & 52238 &  0.77 & 237.3 &  20.4 &  $-22.7$ &   5.1 &   3.6 \\

     022346.42$-$003908.2 & 51817 & 52238 &  1.67 & 157.4 &  19.3 &  $-25.6$ &  11.9 &  15.3 \\

     021938.28$-$002151.4 & 51817 & 52238 &  0.56 & 270.4 &  20.2 &  $-22.1$ &   5.8 &   4.0 \\

     021951.76$-$002108.2 & 51817 & 52238 &  1.61 & 161.1 &  19.6 &  $-25.2$ &   8.9 &   7.0 \\

       022027.54+011401.7 & 51817 & 52238 &  1.33 & 180.8 &  18.5 &  $-25.9$ &  19.2 &  14.7 \\

       022229.99+004837.5 & 51817 & 52238 &  0.62 & 260.6 &  20.2 &  $-22.3$ &   5.7 &   6.5 \\

       022526.15+010124.0 & 51817 & 52238 &  1.87 & 146.5 &  20.2 &  $-24.9$ &   6.3 &   3.7 \\

       022556.34+001345.3 & 51817 & 52238 &  0.71 & 246.5 &  19.2 &  $-23.7$ &  12.6 &  10.3 \\

       022600.00+003234.5 & 51817 & 52238 &  0.79 & 235.6 &  19.9 &  $-23.2$ &   7.6 &   5.4 \\

       022838.62+003320.1 & 51817 & 52238 &  0.77 & 238.1 &  19.6 &  $-23.5$ &   9.9 &   6.5 \\

     025038.68$-$004739.1 & 51816 & 51877 &  1.84 &  21.5 &  19.1 &  $-26.0$ &  15.4 &  17.3 \\

     025030.77$-$000801.8 & 51816 & 51877 &  1.46 &  24.8 &  18.6 &  $-26.0$ &  17.2 &  20.1 \\

       025701.94+010644.6 & 51816 & 51877 &  2.19 &  19.1 &  21.0 &  $-24.5$ &   2.4 &   3.3 \\

       030600.41+010145.4 & 51817 & 51873 &  2.19 &  17.5 &  18.8 &  $-26.7$ &  14.6 &  10.8 \\

     031226.11$-$003708.9 & 51931 & 52254 &  0.62 & 199.2 &  19.1 &  $-23.4$ &  15.3 &   9.2 \\

     031156.44$-$004156.8 & 51931 & 52254 &  0.96 & 165.2 &  19.5 &  $-24.1$ &  12.3 &  12.5 \\

     031131.40$-$002127.4 & 51931 & 52254 &  1.57 & 125.7 &  19.0 &  $-25.7$ &  17.8 &  16.6 \\

       030458.96+000235.7 & 51931 & 52254 &  0.56 & 206.5 &  19.0 &  $-23.3$ &  13.6 &  16.1 \\

       030911.63+002359.0 & 51931 & 52254 &  0.61 & 200.4 &  17.5 &  $-25.0$ &  38.2 &  36.2 \\

       031127.55+005357.3 & 51931 & 52254 &  1.76 & 117.1 &  19.6 &  $-25.5$ &  11.4 &   8.9 \\

       031129.29+005638.6 & 51931 & 52254 &  1.51 & 128.9 &  20.1 &  $-24.6$ &   7.5 &   7.9 \\

     032253.09$-$001121.6 & 51929 & 51821 &  1.88 &  37.5 &  21.0 &  $-24.2$ &   2.9 &   3.8 \\

       032017.86+000647.6 & 51929 & 51821 &  0.88 &  57.4 &  19.7 &  $-23.7$ &   8.8 &  10.5 \\

     031537.33$-$001811.0 & 51929 & 51821 &  1.81 &  38.5 &  20.5 &  $-24.6$ &   4.8 &   5.7 \\

       031544.54+004220.9 & 51929 & 51821 &  1.88 &  37.5 &  19.5 &  $-25.6$ &  10.4 &   7.8 \\

     033202.33$-$003739.0 & 51810 & 51879 &  0.61 &  42.9 &  18.7 &  $-23.8$ &  20.4 &  21.5 \\

     034520.83$-$004842.8 & 51811 & 51885 &  1.33 &  31.8 &  19.4 &  $-25.0$ &  12.1 &   9.6 \\

     034318.37$-$004447.9 & 51811 & 51885 &  1.75 &  26.9 &  21.3 &  $-23.7$ &   2.9 &   1.5 \\

     034345.92$-$004801.4 & 51811 & 51885 &  0.75 &  42.4 &  20.2 &  $-22.8$ &   6.8 &   4.2 \\

       033821.51+003106.4 & 51811 & 51885 &  1.35 &  31.5 &  20.2 &  $-24.2$ &   6.7 &   5.7 \\

       034106.76+004610.0 & 51811 & 51885 &  0.63 &  45.3 &  18.4 &  $-24.1$ &  24.3 &  20.8 \\

       003732.61+144258.0 & 51817 & 51884 &  2.37 &  19.9 &  21.2 &  $-24.5$ &   3.0 &   2.3 \\

       003424.56+142353.7 & 51817 & 51884 &  0.58 &  42.5 &  19.7 &  $-22.7$ &  10.3 &   8.3 \\

       003520.91+143730.2 & 51817 & 51884 &  1.86 &  23.4 &  20.7 &  $-24.5$ &   4.8 &   3.2 \\

       003434.20+134609.8 & 51817 & 51884 &  0.78 &  37.6 &  20.1 &  $-23.0$ &   7.6 &   5.8 \\

       003240.57+143951.9 & 51817 & 51884 &  1.86 &  23.4 &  20.8 &  $-24.3$ &   4.0 &   3.0 \\

       004142.44+153306.9 & 51868 & 51812 &  0.75 &  32.0 &  20.7 &  $-22.3$ &   3.7 &   3.0 \\

       004337.73+160530.0 & 51868 & 51812 &  1.97 &  18.8 &  20.7 &  $-24.6$ &   3.5 &   3.7 \\

     094156.74$-$002434.0 & 52314 & 52027 &  1.51 & 114.3 &  21.0 &  $-23.6$ &   2.6 &   1.6 \\

     093935.07$-$000801.1 & 52314 & 52027 &  0.91 & 150.1 &  20.1 &  $-23.3$ &   6.0 &   5.5 \\

     093622.06$-$004555.4 & 52314 & 52027 &  1.78 & 103.4 &  20.7 &  $-24.3$ &   3.3 &   2.7 \\

     093751.80$-$003007.9 & 52314 & 52027 &  0.75 & 164.2 &  20.0 &  $-23.0$ &   6.7 &   4.6 \\

     093736.74$-$000732.1 & 52314 & 52027 &  1.79 & 102.9 &  19.7 &  $-25.3$ &   8.3 &   5.1 \\

     093616.86$-$000346.9 & 52314 & 52027 &  1.14 & 134.0 &  19.6 &  $-24.4$ &   9.2 &   7.1 \\

     093448.05$-$002723.3 & 52314 & 52027 &  0.99 & 144.5 &  17.9 &  $-25.7$ &  27.0 &  26.8 \\

     093300.12$-$005336.2 & 52314 & 52027 &  1.08 & 137.8 &  21.2 &  $-22.7$ &   2.0 &   2.2 \\

     093226.56$-$005512.2 & 52314 & 52027 &  0.67 & 172.3 &  21.2 &  $-21.5$ &   1.8 &   3.0 \\

     093233.65$-$003441.9 & 52314 & 52027 &  1.84 & 101.2 &  20.4 &  $-24.7$ &   4.1 &   2.8 \\

     093253.32$-$003854.0 & 52314 & 52027 &  1.11 & 135.9 &  18.8 &  $-25.1$ &  14.1 &  16.8 \\

     093210.68$-$001419.5 & 52314 & 52027 &  1.50 & 114.8 &  20.2 &  $-24.5$ &   5.3 &   3.8 \\

     093150.57$-$001935.2 & 52314 & 52027 &  1.84 & 101.1 &  19.6 &  $-25.6$ &   8.8 &   8.3 \\

       093203.21+002608.1 & 52314 & 52027 &  1.02 & 142.2 &  20.0 &  $-23.7$ &   5.8 &   2.8 \\

       093503.28+003001.7 & 52314 & 52027 &  1.47 & 116.2 &  20.2 &  $-24.3$ &   5.2 &   5.1 \\

       093756.64+000508.8 & 52314 & 52027 &  1.08 & 138.3 &  20.7 &  $-23.2$ &   3.9 &   3.8 \\

       093837.25+000439.5 & 52314 & 52027 &  1.49 & 115.1 &  19.5 &  $-25.1$ &  10.3 &   7.5 \\

       093948.10+003446.2 & 52314 & 52027 &  0.65 & 173.7 &  18.5 &  $-24.2$ &  19.5 &  18.2 \\

       094149.60+003254.3 & 52314 & 52027 &  2.00 &  95.6 &  20.9 &  $-24.4$ &   3.1 &   3.5 \\

       094058.03+000344.8 & 52314 & 52027 &  1.52 & 113.9 &  20.6 &  $-24.0$ &   2.4 &   3.3 \\

       131522.44+013917.0 & 52295 & 52029 &  1.67 &  99.8 &  20.8 &  $-24.1$ &   3.5 &   3.1 \\

       131259.37+013022.4 & 52295 & 52029 &  1.55 & 104.3 &  19.3 &  $-25.4$ &  12.3 &  10.6 \\

       131439.23+021214.9 & 52295 & 52029 &  1.78 &  95.8 &  20.2 &  $-24.8$ &   6.3 &   5.7 \\

       131124.01+012643.5 & 52295 & 52029 &  0.51 & 176.5 &  19.0 &  $-23.0$ &  15.0 &  15.8 \\

       131157.99+013517.5 & 52295 & 52029 &  1.08 & 127.9 &  18.6 &  $-25.3$ &  20.1 &  19.3 \\

       130809.22+021203.2 & 52295 & 52029 &  0.91 & 138.9 &  19.0 &  $-24.4$ &  16.2 &   9.6 \\

       130754.44+021820.2 & 52295 & 52029 &  1.87 &  92.7 &  19.9 &  $-25.2$ &   7.9 &   4.4 \\

       130855.25+030614.2 & 52295 & 52029 &  1.79 &  95.5 &  19.8 &  $-25.3$ &   9.8 &   7.7 \\

       130940.60+031826.7 & 52295 & 52029 &  2.76 &  70.8 &  20.3 &  $-25.7$ &   5.8 &   3.3 \\

       130957.69+030452.2 & 52295 & 52029 &  1.47 & 107.5 &  18.7 &  $-25.9$ &  19.9 &  13.8 \\

       130825.64+025736.0 & 52295 & 52029 &  1.75 &  96.6 &  19.5 &  $-25.5$ &  10.9 &   4.2 \\

       131448.31+025603.3 & 52295 & 52029 &  0.77 & 150.5 &  18.9 &  $-24.2$ &  18.0 &  10.9 \\

       082643.25+434648.7 & 52207 & 51959 &  1.48 &  99.9 &  19.8 &  $-24.8$ &   9.7 &   7.7 \\

       081859.78+423327.6 & 52207 & 51959 &  2.22 &  77.0 &  21.2 &  $-24.4$ &   3.7 &   2.3 \\

       081536.62+431455.5 & 52207 & 51959 &  0.55 & 160.1 &  19.1 &  $-23.1$ &  19.6 &  11.8 \\

       081151.35+434730.3 & 52207 & 51959 &  1.64 &  94.1 &  20.3 &  $-24.6$ &   7.8 &   4.5 \\

       081353.65+445159.3 & 52207 & 51959 &  1.58 &  96.1 &  20.7 &  $-24.1$ &   5.9 &   5.4 \\

       081349.01+441517.7 & 52207 & 51959 &  2.21 &  77.1 &  19.6 &  $-26.0$ &  13.6 &  11.4 \\

       081614.97+435640.2 & 52207 & 51959 &  1.96 &  83.8 &  20.8 &  $-24.5$ &   5.2 &   4.8 \\

       081926.51+445759.9 & 52207 & 51959 &  1.71 &  91.4 &  20.0 &  $-25.0$ &   9.3 &   6.8 \\

       082159.11+441858.8 & 52207 & 51959 &  1.45 & 101.0 &  20.5 &  $-24.0$ &   6.5 &   6.3 \\

       082310.94+442048.1 & 52207 & 51959 &  1.78 &  89.2 &  19.6 &  $-25.5$ &  13.9 &  12.6 \\

       082118.08+450637.6 & 52207 & 51959 &  1.31 & 107.3 &  20.1 &  $-24.2$ &   8.5 &   5.9 \\

       153001.69+540452.3 & 52374 & 52442 &  1.72 &  25.0 &  19.0 &  $-25.9$ &  13.8 &   8.7 \\

       160630.61+505408.7 & 52375 & 52081 &  1.25 & 130.9 &  20.7 &  $-23.6$ &   3.8 &   1.5 \\

       155757.23+503546.3 & 52375 & 52081 &  0.74 & 169.1 &  20.1 &  $-22.8$ &   6.8 &   3.7 \\

       160018.14+493720.9 & 52375 & 52081 &  1.16 & 135.8 &  19.1 &  $-25.0$ &  13.5 &   6.0 \\

       160126.31+511038.1 & 52375 & 52081 &  1.84 & 103.4 &  19.8 &  $-25.4$ &   9.4 &   3.7 \\

     231040.97$-$010823.0 & 52884 & 52534 &  2.07 & 114.2 &  19.6 &  $-25.8$ &   8.7 &   7.2 \\

     231202.86$-$005537.6 & 52884 & 52534 &  1.60 & 134.4 &  20.0 &  $-24.8$ &   6.2 &   5.6 \\

     231133.08$-$001449.3 & 52884 & 52534 &  1.38 & 147.2 &  19.5 &  $-24.9$ &   9.5 &   7.0 \\

     230952.29$-$003138.9 & 52884 & 52534 &  3.97 &  70.4 &  20.0 &  $-26.8$ &   6.7 &   4.5 \\

     230728.90$-$011608.9 & 52884 & 52534 &  1.98 & 117.4 &  20.2 &  $-25.1$ &   6.4 &   5.6 \\

     230740.72$-$004948.7 & 52884 & 52534 &  0.69 & 206.5 &  19.9 &  $-22.9$ &   6.7 &   6.3 \\

     230832.98$-$002332.4 & 52884 & 52534 &  3.08 &  85.7 &  21.2 &  $-25.1$ &   2.6 &   1.1 \\

     230653.99$-$001605.5 & 52884 & 52534 &  0.56 & 225.0 &  19.2 &  $-23.0$ &  12.6 &   8.2 \\

     230419.24$-$003237.4 & 52884 & 52534 &  1.22 & 157.4 &  19.9 &  $-24.3$ &   7.3 &   5.4 \\

     230437.65$-$005703.3 & 52884 & 52534 &  2.49 & 100.2 &  20.3 &  $-25.5$ &   5.8 &   4.5 \\

     230402.78$-$003855.4 & 52884 & 52534 &  2.77 &  92.8 &  20.8 &  $-25.3$ &   3.3 &   2.4 \\

     230358.66$-$001733.0 & 52884 & 52534 &  1.51 & 139.4 &  20.5 &  $-24.1$ &   4.6 &   2.5 \\

     230350.33$-$005336.6 & 52884 & 52534 &  0.86 & 188.5 &  20.6 &  $-22.7$ &   4.4 &   3.4 \\

     230424.87$-$010140.8 & 52884 & 52534 &  1.89 & 121.1 &  21.0 &  $-24.2$ &   3.0 &   2.6 \\

       230239.68+002702.5 & 52884 & 52534 &  1.86 & 122.2 &  20.9 &  $-24.2$ &   3.1 &   3.1 \\

       230555.48+005946.2 & 52884 & 52534 &  0.72 & 203.5 &  18.5 &  $-24.4$ &  20.1 &  16.9 \\

       230524.47+005209.7 & 52884 & 52534 &  1.85 & 123.0 &  20.0 &  $-25.1$ &   6.5 &   2.7 \\

       230323.77+001615.1 & 52884 & 52534 &  3.69 &  74.6 &  20.5 &  $-26.1$ &   4.3 &   3.6 \\

       230435.93+003001.5 & 52884 & 52534 &  2.00 & 116.6 &  20.8 &  $-24.6$ &   3.4 &   2.8 \\

       230522.11+001949.2 & 52884 & 52534 &  1.64 & 132.5 &  19.9 &  $-24.9$ &   7.6 &   6.1 \\

       230630.22+001857.5 & 52884 & 52534 &  1.25 & 155.3 &  20.4 &  $-23.8$ &   5.2 &   4.2 \\

       231000.16+003208.0 & 52884 & 52534 &  1.50 & 140.1 &  19.7 &  $-25.0$ &  10.2 &   7.1 \\

       231040.38+000334.5 & 52884 & 52534 &  1.51 & 139.5 &  20.8 &  $-23.9$ &   4.0 &   3.4 \\

       231132.65+003321.2 & 52884 & 52534 &  1.01 & 173.9 &  19.6 &  $-24.1$ &  10.3 &   8.0 \\

       231121.98+004959.7 & 52884 & 52534 &  2.06 & 114.3 &  19.6 &  $-25.8$ &  10.4 &   5.6 \\

       231147.90+002941.9 & 52884 & 52534 &  1.90 & 120.6 &  19.7 &  $-25.5$ &   9.5 &   7.1 \\

       231241.77+002450.3 & 52884 & 52534 &  1.89 & 121.0 &  19.0 &  $-26.2$ &  15.9 &   7.7 \\

       224005.09+143147.8 & 52520 & 52264 &  3.49 &  57.0 &  20.4 &  $-26.1$ &   4.8 &   1.1 \\

       224125.09+143331.2 & 52520 & 52264 &  1.53 & 101.0 &  19.6 &  $-25.1$ &  10.0 &   2.4 \\

       224234.89+145647.7 & 52520 & 52264 &  1.49 & 102.9 &  18.8 &  $-25.9$ &  17.3 &   5.8 \\

       075153.67+331319.8 & 52237 & 52577 &  1.93 & 116.1 &  19.1 &  $-26.1$ &  13.9 &   7.6 \\

       075217.23+335524.5 & 52237 & 52577 &  1.68 & 126.7 &  20.1 &  $-24.8$ &   5.8 &   5.1 \\

       075318.63+335429.8 & 52237 & 52577 &  1.22 & 153.0 &  19.4 &  $-24.7$ &  10.3 &  10.4 \\

       075143.07+331255.6 & 52237 & 52577 &  1.16 & 157.4 &  19.9 &  $-24.2$ &   7.6 &   6.9 \\

       075004.96+334954.6 & 52237 & 52577 &  1.55 & 133.4 &  18.6 &  $-26.2$ &  20.7 &  19.0 \\

       074823.86+332051.2 & 52237 & 52577 &  2.99 &  85.2 &  20.0 &  $-26.2$ &   6.3 &   9.1 \\

       074915.29+343859.3 & 52237 & 52577 &  0.87 & 181.8 &  18.2 &  $-25.2$ &  25.9 &  28.8 \\

       075132.75+350535.0 & 52237 & 52577 &  2.07 & 110.9 &  20.7 &  $-24.7$ &   4.3 &   2.8 \\

       075300.91+350821.0 & 52237 & 52577 &  1.55 & 133.1 &  19.6 &  $-25.1$ &  10.1 &  12.3 \\

       075321.93+350733.5 & 52237 & 52577 &  1.90 & 117.4 &  20.7 &  $-24.5$ &   3.8 &   5.0 \\

       075826.26+345019.9 & 52237 & 52577 &  0.88 & 180.7 &  20.5 &  $-22.9$ &   5.2 &   3.1 \\

       075614.59+350414.4 & 52237 & 52577 &  0.83 & 185.3 &  19.9 &  $-23.3$ &   8.2 &   5.1 \\

       075715.75+345424.4 & 52237 & 52577 &  0.83 & 185.4 &  18.6 &  $-24.7$ &  21.4 &  15.8 \\

       144059.16+573724.3 & 52346 & 52433 &  2.04 &  28.6 &  19.9 &  $-25.5$ &   5.9 &   5.7 \\

       143905.75+574523.4 & 52346 & 52433 &  1.62 &  33.2 &  20.3 &  $-24.5$ &   5.2 &   3.2 \\

       143618.60+581044.2 & 52346 & 52433 &  1.65 &  32.8 &  20.8 &  $-24.1$ &   3.9 &   2.3 \\

       143556.71+581522.5 & 52346 & 52433 &  1.62 &  33.2 &  20.5 &  $-24.3$ &   4.7 &   2.5 \\

       145316.61+560750.8 & 52347 & 52435 &  1.85 &  30.9 &  20.9 &  $-24.2$ &   2.7 &   2.5 \\

       144627.87+563836.5 & 52347 & 52435 &  0.62 &  54.2 &  20.3 &  $-22.2$ &   5.0 &   4.4 \\

       144047.43+562910.9 & 52347 & 52435 &  0.70 &  51.8 &  21.1 &  $-21.7$ &   2.7 &   2.2 \\

       143632.31+563319.5 & 52347 & 52435 &  1.77 &  31.8 &  19.4 &  $-25.6$ &   8.9 &   7.0 \\

       144106.82+574939.9 & 52347 & 52435 &  1.23 &  39.5 &  18.2 &  $-26.0$ &  20.6 &  22.3 \\

       161758.81+442259.4 & 52443 & 52355 &  0.58 &  55.9 &  19.2 &  $-23.1$ &  11.7 &   4.7 \\

       161128.81+444142.5 & 52443 & 52355 &  1.08 &  42.4 &  19.3 &  $-24.5$ &  11.3 &   4.2 \\

       161240.98+435749.4 & 52443 & 52355 &  1.74 &  32.1 &  20.6 &  $-24.3$ &   4.0 &   0.7 \\

       102003.81+474019.2 & 52347 & 52674 &  1.03 & 161.4 &  19.0 &  $-24.8$ &  12.1 &  11.0 \\

       102248.34+475547.4 & 52347 & 52674 &  0.74 & 187.7 &  19.1 &  $-23.8$ &  11.3 &   7.6 \\

       101754.86+470529.3 & 52347 & 52674 &  0.67 & 196.1 &  18.6 &  $-24.1$ &  14.6 &  13.0 \\

       101821.61+473251.4 & 52347 & 52674 &  1.51 & 130.3 &  18.2 &  $-26.4$ &  19.3 &  16.5 \\

       101902.02+473714.5 & 52347 & 52674 &  2.95 &  82.8 &  19.2 &  $-26.9$ &   9.7 &   7.8 \\

       101621.95+474908.2 & 52347 & 52674 &  1.35 & 139.3 &  19.1 &  $-25.3$ &  10.6 &  10.0 \\

       101728.77+481331.5 & 52347 & 52674 &  1.63 & 124.5 &  19.2 &  $-25.6$ &   9.9 &  11.8 \\

       102048.82+483908.8 & 52347 & 52674 &  1.94 & 111.2 &  18.7 &  $-26.5$ &  17.5 &  16.9 \\

       102532.69+483039.3 & 52347 & 52674 &  0.59 & 205.2 &  18.7 &  $-23.7$ &  15.7 &  14.7 \\

       102400.88+492359.3 & 52347 & 52674 &  0.54 & 212.8 &  19.4 &  $-22.8$ &   9.6 &   9.7 \\

       105922.46+494918.2 & 52346 & 52669 &  1.68 & 120.5 &  20.4 &  $-24.5$ &   2.7 &   5.9 \\

       105430.08+491947.1 & 52346 & 52669 &  4.00 &  64.6 &  19.8 &  $-27.1$ &   4.8 &   8.5 \\

       105555.57+501745.6 & 52346 & 52669 &  0.69 & 191.6 &  20.2 &  $-22.5$ &   4.8 &   4.2 \\

       105027.74+490453.0 & 52346 & 52669 &  1.86 & 112.8 &  19.6 &  $-25.6$ &   6.2 &  11.1 \\

       104951.09+493156.2 & 52346 & 52669 &  1.79 & 115.7 &  19.7 &  $-25.3$ &   7.4 &   9.6 \\

       105038.30+500411.9 & 52346 & 52669 &  0.68 & 192.4 &  19.8 &  $-22.9$ &   7.6 &   8.2 \\

       104806.47+501021.5 & 52346 & 52669 &  1.78 & 116.0 &  20.9 &  $-24.2$ &   3.8 &   3.6 \\

       104859.93+504715.2 & 52346 & 52669 &  1.00 & 161.5 &  18.6 &  $-25.0$ &  21.0 &  13.5 \\

       105421.34+514132.8 & 52346 & 52669 &  1.11 & 153.3 &  18.8 &  $-25.1$ &  18.4 &  12.7 \\

       105454.16+503123.9 & 52346 & 52669 &  1.87 & 112.4 &  18.8 &  $-26.3$ &  19.1 &  12.4 \\

       105534.06+502624.7 & 52346 & 52669 &  0.81 & 178.6 &  20.8 &  $-22.4$ &   3.5 &   2.4 \\

       105457.60+510958.4 & 52346 & 52669 &  0.73 & 187.1 &  19.7 &  $-23.2$ &   9.6 &   7.6 \\

       105657.90+510232.5 & 52346 & 52669 &  1.37 & 136.3 &  18.5 &  $-25.9$ &  22.3 &  14.1 \\

       074641.88+291904.5 & 52346 & 52663 &  0.80 & 175.7 &  19.1 &  $-24.1$ &  15.5 &  10.4 \\

       074641.95+293247.9 & 52346 & 52663 &  2.28 &  96.8 &  19.4 &  $-26.2$ &  11.8 &  10.0 \\

       074621.32+292821.8 & 52346 & 52663 &  1.43 & 130.3 &  19.1 &  $-25.5$ &  15.4 &  13.2 \\

       074451.37+292005.9 & 52346 & 52663 &  1.18 & 145.2 &  16.8 &  $-27.3$ &  53.4 &  41.8 \\

       074321.71+283840.8 & 52346 & 52663 &  1.16 & 146.5 &  18.7 &  $-25.4$ &  22.7 &  16.5 \\

       074317.90+290622.4 & 52346 & 52663 &  0.71 & 185.8 &  19.1 &  $-23.8$ &  17.3 &  12.0 \\

       074407.41+294707.4 & 52346 & 52663 &  1.86 & 110.8 &  19.4 &  $-25.8$ &  10.9 &  14.0 \\

       074227.83+290000.4 & 52346 & 52663 &  1.13 & 148.6 &  19.1 &  $-24.9$ &  17.2 &  12.8 \\

       074209.63+293657.2 & 52346 & 52663 &  1.57 & 123.4 &  19.5 &  $-25.3$ &  11.5 &   8.9 \\

       074311.16+303214.9 & 52346 & 52663 &  1.64 & 120.1 &  19.2 &  $-25.7$ &  18.2 &  12.5 \\

       074600.91+303124.4 & 52346 & 52663 &  0.82 & 174.3 &  19.3 &  $-23.9$ &  15.6 &  13.6 \\

       074625.28+302020.7 & 52346 & 52663 &  1.74 & 115.9 &  18.4 &  $-26.6$ &  27.8 &  21.7 \\

       074635.06+295645.6 & 52346 & 52663 &  0.90 & 167.0 &  19.5 &  $-23.9$ &  13.0 &   9.6 \\

       074937.74+304021.4 & 52346 & 52663 &  1.73 & 116.2 &  20.9 &  $-24.1$ &   3.7 &   4.8 \\

       082443.39+055503.7 & 52962 & 52737 &  2.10 &  72.5 &  18.9 &  $-26.5$ &  13.2 &   9.0 \\

       082328.61+061146.0 & 52962 & 52737 &  2.78 &  59.5 &  18.1 &  $-27.9$ &  24.9 &  14.6 \\

       082256.01+060528.7 & 52962 & 52737 &  1.98 &  75.4 &  19.6 &  $-25.7$ &   9.4 &   4.2 \\

       082216.57+060344.9 & 52962 & 52737 &  1.58 &  87.0 &  19.4 &  $-25.3$ &  10.8 &   5.4 \\

       082202.31+061340.0 & 52962 & 52737 &  0.82 & 123.4 &  19.6 &  $-23.6$ &   9.3 &   5.0 \\

       081941.12+054942.6 & 52962 & 52737 &  1.70 &  83.3 &  21.0 &  $-24.0$ &   2.8 &   2.3 \\

       081931.48+055523.6 & 52962 & 52737 &  1.69 &  83.7 &  18.3 &  $-26.6$ &  21.1 &  12.6 \\

       081811.50+053713.9 & 52962 & 52737 &  2.51 &  64.1 &  18.6 &  $-27.2$ &  17.7 &  11.2 \\

       082257.04+070104.3 & 52962 & 52737 &  2.95 &  56.9 &  18.7 &  $-27.5$ &  17.5 &  10.2 \\

       081941.33+064036.9 & 52962 & 52737 &  1.63 &  85.4 &  19.8 &  $-25.1$ &   7.4 &   3.1 \\

       082337.40+064436.0 & 52962 & 52737 &  1.02 & 111.4 &  18.5 &  $-25.2$ &  19.3 &  17.5 \\

       082647.18+065406.0 & 52962 & 52737 &  0.97 & 114.5 &  18.0 &  $-25.6$ &  25.5 &  16.3 \\
\enddata
\end{deluxetable}

\end{document}